\documentclass[10pt, a4paper]{article}
\usepackage{jheppub}
\usepackage{bm}
\usepackage{mathrsfs}
\usepackage{dsfont}
\usepackage{slashed}
\usepackage{braket}
\usepackage{float}
\usepackage{mathtools}

\usepackage{indentfirst}
\usepackage{caption, subcaption}
\usepackage{physics}
\usepackage{xcolor}
\usepackage{multirow}
\usepackage{bbding}
\usepackage{youngtab}
\usepackage{young}
\usepackage{ytableau}
\usepackage{bbm}
\usepackage[utf8]{inputenc}
\usepackage{array}
\usepackage{colortbl}
\usepackage{hhline}
\usepackage{longtable}
\usepackage{arydshln}
\usepackage{tabularx}
\usepackage{makecell} 
\usepackage{ragged2e} 
\usepackage{booktabs} 

\newcommand{\onec}{1^{c}}
\newcommand{\twoc}{2^{c}}
\newcommand{\threec}{3^{c}}
\newcommand{\fourc}{4^{c}}
\newcommand{\fivec}{5^{c}}
\newcommand{\sixc}{6^{c}}

\newcommand{\blue}[1]{\textcolor{blue}{#1}}

\newcommand{\antiket}[1]{| #1 ]}

\newcommand{\sbk}[1]{\left[{#1}\right]}
\newcommand{\abk}[1]{\left\langle{#1}\right\rangle}

\newcommand{\pd}{d}
\allowdisplaybreaks[4] 
\newcommand{\be}{\begin{equation}}
\newcommand{\ee}{\end{equation}}
\newcommand{\bea}{\begin{eqnarray}}
\newcommand{\eea}{\end{eqnarray}}
\newcommand{\mOne}{\mathbf{1}}
\newcommand{\mTwo}{\mathbf{2}}

\title{Dark Photons and Gravitino Like Particles: Complete EFT Operator Basis}

\author[a]{Ziyu Dong}
\author[b]{Teng Ma}
\author[b]{Chengjie Yang}
\author[c, d]{Zizheng Zhou}

\affiliation[a]{IFAE and BIST, Universitat Aut\`onoma de Barcelona, 08193 Bellaterra, Barcelona}
\affiliation[b]{International Center for Theoretical Physics Asia-Pacific (ICTP-AP),\\ University of Chinese Academy of Sciences, 100190 Beijing, China}
\affiliation[c]{CAS Key Laboratory of Theoretical Physics, Institute of Theoretical Physics, \\Chinese Academy of Sciences, Beijing 100190, China}
\affiliation[d]{School of Physical Sciences, University of Chinese Academy of Sciences, Beijing 100049, China}

\emailAdd{zdong@ifae.es}
\emailAdd{mateng@ucas.ac.cn}
\emailAdd{yangchengjie@ucas.ac.cn}
\emailAdd{zhouzizheng@itp.ac.cn}

\date{Version: Draft built on \today}

\abstract{We present a more efficient method for constructing the complete EFT operator basis for particles of any mass and spin, based on on-shell method and Young tableaux. By classifying the amplitude bases according to the polarization configurations of massive particles and using their high-energy limit, our approach can construct EFT basis in a straightforward way, without need of auxiliary fields and tedious basis decomposition. Based on this improved method, we develop a \textit{Mathematica} code that can automatically construct the EFT basis for particles of any spin. As applications, the EFT bases up to \( d = 8 \) are explicitly constructed for dark photons and, for the first time, for spin-\( 3/2 \) gravitino like particles.
}

\begin{document}

\maketitle
\flushbottom

\tableofcontents

\section{Introduction}

EFT has become an indispensable tool in dark matter research, providing a model-independent framework to characterize interactions between dark matter and Standard Model (SM) particles. Moreover, EFT is widely applied in precise measurements due to its simplicity in loop calculations. However, the complete operator basis is critical for consistent EFT calculations, such as calculating the renormalization running. Therefore, constructing the complete EFT basis is essential for new physics searches at lower energy scales. But directly constructing the operators is very difficult because there are two kinds of redundancies that need to be eliminated: integration by parts (IBP) and equations of motion (EOM). So far, there is no systematic way to deal with them in traditional field theories. Based on the on-shell scattering amplitude method, it was found that the independent polynomials of the product of the on-shell helicity spinor variables correspond one-to-one to the independent EFT operators. For the on-shell EFT bases of massless fields, there is only IBP redundancy, which can be systematically removed using $SU(N)$ Young tableaux, where $N$ is the number of external legs~\cite{Henning:2019enq}. However, for the on-shell bases involving massive fields, this method cannot be applied anymore. To solve this, a series of works~\cite{Dong:2022mcv,Dong:2021vxo} established a general framework to construct the on-shell EFT bases of massive fields with any spin, based on Lorentz symmetry and the Young tableau method. Using this method, the bases of Higgs effective field theory at higher dimensions can be automatically constructed by programs~\cite{Dong:2022jru} (other alternative approaches for constructing massive EFT can be found in~\cite{Balkin:2021dko,Goldberg:2024eot,Christensen:2024bdt,Graf:2022rco,DeAngelis:2022qco}).
   
In recent years, since high-scale new physics (NP) has not been found at the LHC~\cite{Alekhin:2015byh,Arcadi:2017kky,Beacham:2019nyx,CMS:2014jvv,LHCNewPhysicsWorkingGroup:2011mji,ParticleDataGroup:2018ovx,Contardo:2015bmq,Abercrombie:2015wmb,Abdallah:2015ter,Alimena:2019zri,Drees:2013wra,Buchmueller:2013dya,Buckley:2014fba,CMS:2013wdg} and other experiments~\cite{Guo:2019qgs,PPTA:2021uzb,Guo:2020drq,Chen:2021bdr,Antypas:2022asj,Jiang:2015cwa,Chao:2017vrq,Caldwell:2022qsj,Murayama:2009nj}, new physics with a mass scale much lighter than the electroweak breaking scale has become increasingly popular, such as axion-like particles (ALPs), dark photons, and higher spin resonances.
Since the masses of dark photons or ALPs are protected by gauge symmetry or shift symmetry, their lower mass scale is technically natural. Generally, their UV completion may contain other resonances much heavier than the electroweak symmetry breaking scale. Therefore, the UV theory at scales much lower than the UV scale but above the electroweak scale can be described by an EFT of massless SM fields and the light NP. This kind of EFT is widely applied in the search for light NP at lower energy scales. The EFT operators are critical in predicting experimental signatures, such as missing energy or displaced vertices in high-energy colliders, and direct detection signals from interactions with ordinary matter.

In this work, we focus on the EFT of the dark photon. The dark photon is present in many new physics models, such as the Twin Higgs model~\cite{Chacko:2005pe,Bellazzini:2014yua,Carmi:2012in,Chacko:2005un,Low:2015nqa,Chang:2006ra,Barbieri:2015lqa,GarciaGarcia:2015fol,Chacko:2016hvu,Csaki:2019qgb,Burdman:2006tz,Craig:2014aea,Csaki:2017jby,Geller:2014kta,Serra:2017poj,Dong:2020eqy} and various dark matter models~\cite{Arias:2012az,Delaunay:2020vdb}. This kind of field, associated with a gauge symmetry $U(1)_D$, can interact with SM fields in different ways, such as through kinetic mixing~\cite{Fabbrichesi:2020wbt,Cheung:2009qd,Gherghetta:2019coi,Goodsell:2011wn,Choi:2013qra,Hapitas:2021ilr} or by directly coupling to SM fields~\cite{Fayet:2016nyc,Bauer:2018onh,Heeck:2014zfa}. The dark photon can act as a mediator between the dark sector and the SM sector or serve as a dark matter candidate.           

A more efficient approach is introduced for constructing complete EFT operators for particles of any mass and spin.
By classifying amplitude basis based on polarization configurations and utilizing their high-energy limit, our approach eliminates the need for auxiliary fields~\cite{Song:2023jqm,Song:2023lxf} or tedious decompositions~\cite{Dong:2021vxo,Dong:2022mcv,Dong:2022jru}.
In order to demonstrate the advantages of this method, we also construct the EFT basis for fields with higher spin, such as spin $= 3/2$, which can play the role of the graviton's supersymmetry partner~\cite{Kitazawa:2018zys,Dudas:2000nv,Sugimoto:1999tx,Antoniadis:1999xk,Angelantonj:1999jh,Aldazabal:1999jr} or dark matter candidate~\cite{liao2012JHEP, yang2020PRD}. Their complete bases up to $\text{d}=8$ have not been constructed so far, so these bases are useful for phenomenological studies.

The structure of the paper is as follows. Section \ref{sec:Correspondence} describes the on-shell method and the correspondence between EFT operator and amplitude basis. Section \ref{sec:amplitude-basis} introduces our approach's framework in constructing the massive amplitude basis, gauge factor and the identical particles. Section \ref{sec:construction-basis} discusses our approach in detail. Section \ref{sec:example} gives an example to demonstrate our method. We reach our conclusion in Section \ref{sec:conclusion}.
The conventions for the basis expressions and details of constructing bases involving identical particles are presented in Appendix~\ref{sec:convention} and Appendix~\ref{sec:identical}.
The complete EFT operators up to dimension 8 for dark photon and gravitino like particles are presented in Appendix~\ref{app:DP} and \ref{app:Gravitino}, respectively.
\section{Correspondence between EFT Operators and Amplitude Bases}
\label{sec:Correspondence}

EFT operators describe local interactions in the Lagrangian as contact terms without additional singularities. The low-energy Lagrangian is expressed as:
\begin{equation}
\label{eq:eft_lagrangian}
\mathcal{L}_{\text{low-energy}} = \mathcal{L}_{\text{SM}} + \sum \frac{\mathcal{O}_d}{\Lambda^{d-4}},
\end{equation}
where $\mathcal{O}_d$ represents EFT operators of dimension $d$, and $\Lambda$ is the cutoff scale. Each independent operator corresponds to an on-shell scattering amplitude~\cite{Shadmi:2018xan,Harlander:2023ozs,Ma:2019gtx,Elvang:2010jv}, with associated Wilson coefficients constrained by experiments, bootstrap methods~\cite{Arkani-Hamed:2020blm,deRham:2017avq,Caron-Huot:2020cmc,Bern:2021ppb,Chiang:2021ziz,Davighi:2021osh,Du:2021byy,EliasMiro:2021nul,Tolley:2020gtv,Bellazzini:2020cot,Chowdhury:2021ynh,Fernandez:2022kzi,Ma:2023vgc,Dong:2024omo}, and astronomical observations.

For the dark photon, the $U(1)_D$ gauge symmetry breaking scale is often assumed to be significantly higher than the electroweak symmetry breaking (EWSB) scale~\cite{Fabbrichesi:2020wbt}. If the dark photon is lighter than the EWSB scale, its low-energy effects can be described by an EFT involving massive dark photons and SM fields in the pre-EWSB phase.
This setup is highly applicable to  phenomenological studies. In the rest of this work, we use on-shell methods to construct such EFTs for dark photons, starting with a review of the basic properties of on-shell amplitudes.

The on-shell amplitudes for massless particles can be expressed using spinor-helicity variables $\lambda_\alpha$ and $\tilde{\lambda}_{\dot\alpha}$, derived from the decomposition of the on-shell massless momentum matrix:
\begin{equation}
p_{\dot\alpha \alpha} \equiv p_\mu \sigma^\mu_{\dot\alpha \alpha} = \tilde{\lambda}_{\dot\alpha} \lambda_\alpha,
\end{equation}
where $\sigma^\mu$ are the four-dimensional Pauli matrices. The left-handed spinors $\lambda$ and right-handed spinors $\tilde{\lambda}$ transform as doublets under the Lorentz subgroups $SU(2)_l$ and $SU(2)_r$ (with $SO(3,1) \cong SU(2)_l \times SU(2)_r$) and carry $\pm 2h$ units of little group (LG) $U(1)$ charge, where $h$ is the helicity~\cite{Witten:2003nn}.

For massive particles, the spinors $\lambda^I_\alpha$ and $\tilde{\lambda}^I_{\dot\alpha}$ are obtained from the massive momentum matrix, where the superscript $I$ denotes the index of fundamental representation of the massive LG $SU(2)$. Similar to the Dirac equation, the corresponding EOMs are:
\begin{equation}
p_i^\mu \sigma_\mu |i^I] = m_i |i^I\rangle\,, \quad p_i^\mu \sigma_\mu |i^I\rangle = m_i |i^I],
\end{equation}
with the notations $|i^I] \equiv \tilde{\lambda}_i^I$ and $|i^I\rangle \equiv \lambda^I_i$.

Minimal Lorentz-invariant variables are formed by pairs of left-handed or right-handed spinors:
\begin{equation}
[ij]^{IJ} \equiv \epsilon^{\dot\alpha \dot\beta} |i^I]_{\dot\alpha} |j^J]_{\dot\beta}, \quad
\langle ij\rangle^{IJ} \equiv \epsilon^{\alpha \beta} \langle i^I|_\alpha \langle j^J|_\beta,
\end{equation}
known as square and angle brackets. Scattering amplitudes can thus be expressed as functions of these spinor brackets.

Since the EFT operator basis corresponds to independent local interactions, the amplitude basis must consist of independent spinor polynomials constrained by LG symmetry, gauge symmetry, and spin statistics. The correspondence between the EFT operators and the scattering amplitude bases is summarized in Tables~\ref{tab:sm_buildingblocks} and~\ref{tab:dm_buildingblocks}. Using these correspondences, on-shell amplitude bases can be directly translated into EFT operators. Table~\ref{tab:sm_buildingblocks} maps massless SM fields to helicity spinor variables, while Table~\ref{tab:dm_buildingblocks} shows the correspondence for the dark photon $X_\mu$ and spin-3/2 fermion $\psi_\mu$. The arrow $\Leftrightarrow$ denotes that different polarizations interconvert via EOMs. Techniques for constructing these bases using Young tableaux will be detailed in the following sections.
\begin{table}[htbp]
    \centering
    \renewcommand{\arraystretch}{1.5} 
    \setlength{\tabcolsep}{10pt} 
    \begin{tabular}{c|c|c|c|c|c}
    \hline
    SM Particles & Fields & Spinors & $SU(3)_c$ & $SU(2)_L$ & $U(1)_Y$ \\
    \hline
    \multirow{3}{*}{Gauge boson}
    & $G^L_{\mu\nu}\,\sigma^{\mu\nu}$ & $\lambda\lambda$ & $\vcenter{\hbox{\scalebox{0.6}{\yng(2,1)}}}$ & $\mathds{1}$ & $0$ \\ 
    \cline{2-6}
    & $W^L_{\mu\nu}\sigma^{\mu\nu}$ & $\lambda\lambda$ & $\mathds{1}$ & $\vcenter{\hbox{\scalebox{0.6}{\yng(2)}}}$ & $0$ \\
    \cline{2-6}
    & $B^L_{\mu\nu}\sigma^{\mu\nu}$ & $\lambda\lambda$ & $\mathds{1}$ & $\mathds{1}$ & $0$ \\
    \hline
    \multirow{6}{*}{Fermion}
    & $L$ & $\lambda$ & $\mathds{1}$ & $\vcenter{\hbox{\scalebox{0.6}{\yng(1)}}}$ & $-\frac{1}{2}$ \\
    \cline{2-6}
    & $e_R$ & $\tilde{\lambda}$ & $\mathds{1}$ & $\mathds{1}$ & $-1$ \\
    \cline{2-6}
    & $\nu_R$ & $\tilde{\lambda}$ & $\mathds{1}$ & $\mathds{1}$ & $0$ \\
    \cline{2-6}
    & $Q$ & $\lambda$ & $\vcenter{\hbox{\scalebox{0.6}{\yng(1)}}}$ & $\vcenter{\hbox{\scalebox{0.6}{\yng(1)}}}$ & $+\frac{1}{6}$ \\
    \cline{2-6}
    & $u_R$ & $\tilde{\lambda}$ & $\vcenter{\hbox{\scalebox{0.6}{\yng(1)}}}$ & $\mathds{1}$ & $+\frac{2}{3}$ \\
    \cline{2-6}
    & $d_R$ & $\tilde{\lambda}$ & $\vcenter{\hbox{\scalebox{0.6}{\yng(1)}}}$ & $\mathds{1}$ & $-\frac{1}{3}$ \\
    \hline
    \multirow{1}{*}{Higgs}
    & $H$ & $\mathds{1}$ & $\mathds{1}$ & $\vcenter{\hbox{\scalebox{0.6}{\yng(1)}}}$ & $+\frac{1}{2}$ \\
    \hline
    \hline
    Momentum
    & $D_\mu \sigma^\mu$ & $\lambda\tilde{\lambda}$ & $\mathds{1}$ & $\mathds{1}$ & $0$ \\
    \hline
    \end{tabular}
    \caption{Correspondence between fields in EFT operators and spinor-helicity variables in amplitudes for massless SM particles. For simplification we omit the Lorentz indices of $\lambda$ and $\tilde{\lambda}$. Complex conjugates of the fields, which flip helicity, are omitted. Here, $F^{L/R}_{i \mu\nu} \equiv \frac{1}{2}(F_{i\mu \nu} \pm \frac{i}{2} \epsilon_{\mu \nu \rho \sigma} F^{\rho \sigma}_i)$.}
    \label{tab:sm_buildingblocks}
\end{table}

\begin{table}[htbp]
    \centering
    \renewcommand{\arraystretch}{2}
    \begin{tabular}{c|c|c}
    \hline
    New Light Particles & Fields & Spinors \\
    \hline
    Dark photon & $X^L_{\mu\nu}\sigma^{\mu\nu} \Leftrightarrow m X_{\mu}\sigma^\mu \Leftrightarrow X^R_{\mu\nu}\sigma^{\mu\nu}$ & $\lambda^{\{I}\lambda^{J\}} \Leftrightarrow \lambda^{\{I}\tilde{\lambda}^{J\}} \Leftrightarrow \tilde{\lambda}^{\{I}\tilde{\lambda}^{J\}}$ \\
    \hline
    Gravitino like particles& $\begin{aligned}&
        D_\nu\psi^L_\mu\sigma^{\mu\nu} \Leftrightarrow m \psi^L_{\mu}\sigma^\mu \Leftrightarrow \\&m \psi^R_{\mu}\sigma^\mu \Leftrightarrow D_\nu\psi^R_\mu\sigma^{\mu\nu}\end{aligned} $ & $\begin{aligned}&\lambda^{\{I}\lambda^{J}\lambda^{K\}} \Leftrightarrow \lambda^{\{I}\lambda^{J}\tilde{\lambda}^{K\}} \Leftrightarrow \\&\lambda^{\{I}\tilde{\lambda}^{J}\tilde{\lambda}^{K\}} \Leftrightarrow \tilde{\lambda}^{\{I}\tilde{\lambda}^{J}\tilde{\lambda}^{K\}}\end{aligned}$ \\
    \hline
    \end{tabular}
    \caption{Correspondence between fields in EFT operators and polarizations in amplitudes for dark matter candidates. All fields are neutral under the SM gauge group $SU(3)_c \times SU(2)_L \times U(1)_Y$. The $\{\dots\}$ around LG indices indicates symmetrization. $\psi^{L,R}_\mu \equiv P_{L,R} \psi_\mu$ denotes left- or right-handed chirality.}
    \label{tab:dm_buildingblocks}
\end{table}

\section{Amplitude Basis Involving Massive Fields}\label{sec:amplitude-basis}
For EFT operators involving a massive particle-$i$ with spin $s_i$, the corresponding amplitude basis is in the $(2s_i +1)$-dimensional symmetric representation of LG $SU(2)_i$. The transformation is given by:
\begin{equation}
\mathcal{M}^{\{I_{1},\dots, I_{2s_i}\}}\left(w_{I'_1}^{i\,I_1} |i^{I'_1}],\,\dots,\,w_{I'_{2s_i}}^{i\,I_{2s_i}} |i^{I'_{2s_i}}\rangle\right) = w_{I'_1}^{i\,I_1} \cdots w_{I'_{2s_i}}^{i\,I_{2s_i}} \mathcal{M}^{\{I'_1,\dots, I'_{2s_i}\}}\left(|i^{I'_1}],\,\dots,\,|i^{I'_{2s_i}}\rangle\right) \,,
\label{fullD}
\end{equation}
where $w^i$ is the transformation of LG $SU(2)_i$, and the bracket $\{\dots\}$ means the LG indices within it are totally symmetric.

The exposed LG indices arise from the polarization tensor $\varepsilon_i$ of the massive particle. The $\varepsilon_i$ can be expressed in terms of helicity spinor variables and has $(2s_i +1)$ equivalent expressions, denoted as $\varepsilon_i^{l_i}$,
\begin{equation}
\label{eq:polarizationConfig}
\bm{\varepsilon}_i^{l_i} \equiv \left( |i^{\{I}\rangle \right)^{l_i} \left(|i^{I\}}]\right)^{2s_i-l_i}, \quad l_i \in [0,\, 2s_i],
\end{equation}
where $l_i$ denotes the number of left-handed spinors in the massive particle's polarization tensor, labeling the polarization configurations for massive particles. These equivalent expressions  can be transformed into one another via EOM.  As an example, a massive vector has 3 equivalent polarization configurations $\bm{\varepsilon}_i^{l}$, which are $\bm{\varepsilon}_i^{l=0} = |i^{\{I}]|i^{J\}}]$, $\bm{\varepsilon}_i^{l=1} = \ket{i^{\{I}}|i^{J\}}]$, $\bm{\varepsilon}_i^{l=2} = \ket{i^{\{I}}\ket{i^{J\}}}$. For massive particle with spin $\leq 1/2$, its  polarization configurations correspond one-to-one to its helicity states at the high energy limit. However, for the massive particle with spin-$s\geq 1$, the number of its polarization configurations does not match its helicity states in the  massless limit, $h=\pm s$. The degrees of freedom mismatch between UV and IR scale amplitudes     prevents us from constructing a massive amplitude basis through the massless amplitude basis. For example,  
the longitudinal polarization configuration $|i^{\{I}]\langle i^{J\}}|$ of a massive vector and the momentum $|i^{I}]\langle i_{I}|$   
share the same massless limit, so there is ambiguity in the mapping between the UV and IR scattering amplitude basis. In the following subsection we will discuss how to overcome this obstacle.

\subsection{Framework for constructing massive amplitude basis}

Generally one can categorize the amplitude basis $\mathcal{M}$ according to the polarization configurations in Eq.~(\ref{eq:polarizationConfig}). Each category is labeled by the indices $\{l_i\} \equiv \{l_1, l_2, \dots\}$, representing the polarization configurations of all massive particles. The basis in each category can be factorized into two parts, $\mathcal{C}$ which contain all the right-handed spinors from the massive polarization tensor $\varepsilon_i^{l_i}$, and the remaining components denoted as $F$.
The factorization can be expressed as follows:
\begin{equation}
\mathcal{M}^{\{l_i\}}= 
 \mathcal{C}^{\{l_i\}}\left( |i^{\{I}]^{2s_i -l_i} \right)\cdot
F^{\{l_i\}}\left( |i^{I\}}\rangle^{l_i},\, \bm{p}_i,\, \varepsilon_j^{h_j},\, p_j \right),
\label{CFget}
\end{equation}
where indices \(i\) and \(j\) refer to massive and massless particles, respectively, and $\varepsilon_j^{h_j}$ is the polarization tensor with helicity $h_j$ of massless particle. With this factorization, we find that by separately taking the massless limits $F \to f$ and $\mathcal{C} \to c$ and then contracting the resulting massless tensors  we obtain an independent and complete basis for massless fields.
Notice that massless limits for $C$ and $F$ just mean stripping the  LG indices of the massive spinors in $C$ and $F$.     
Conversely, if one first constructs the massless basis $\bigcup_{\{l_i\}} \, \{c \cdot f\}^{\{l_i\}} $,  the independent and redundancy-free massive amplitude basis   
$\mathcal{M}^{\{l_i\}}$ can be obtained by restoring the LG indices of the stripped spinors in the massless limit basis $\{c \cdot f\}$, by mapping the stripped spinors $|i]/|i\rangle$ back to their massive counterparts $|i^I]/|i^I\rangle$. 
This process is referred to as ``{\bf massification}". The relation between massless and massive bases is illustrated as
 \begin{equation}
\{\mathcal{M}\} \xrightleftharpoons[\rm Massification]{\rm Massless\; limit} \bigcup_{\{l_i\}} \, \{c \cdot f\}^{\{l_i\}} \,.
\end{equation}
Notice that the massification mapping is not unique, as different choices for restoring LG indices lead to different massive amplitudes. However, as found in~\cite{Dong:2021vxo,Dong:2022mcv}, since these different amplitudes share the same massless limit, they are related by EOM. So to obtain the independent $\mathcal{M}$ bases, one massification can be chosen arbitrarily. The independence of the massive amplitude bases obtained in this way will be discussed in the next section. The framework for constructing the massive bases can be summarized as follows: first construct the massless limit bases $\bigcup_{\{l_i\}} \, \{c \cdot f\}^{\{l_i\}}$ which are classified by $\{l_i\}$; then choose a massification way to obtain the massive bases $\{\mathcal{M}^{\{l_i\} }\}$; finally the resulting $\{\mathcal{M}^{\{l_i\} }\}$ basis is complete and independent without EOM and IBP redundancies.

\subsection{Gauge factors and exchange symmetries}
Above discussions only focus on the independent Lorentz structure of EFT amplitude bases. The valid amplitude basis should be singlet under the SM gauge symmetry $SU(3)_c \times SU(2)_L \times U(1)_Y$, so gauge factor $\{T\}$ for each basis should be introduced. The $\{T\}$ basis can be systematically constructed via Young diagram method, which is discussed intensively in~\cite{Dong:2022jru}.  

When dealing with identical particles, the requirement of invariance under symmetric permutations imposes additional constraints on the bases $\{T\} \times \{\mathcal{M}\}$. We propose a new perspective: \textit{identical particles with different polarization configurations can be treated as distinct, so exchange symmetries apply only to identical particles with the same polarization configurations.}
To construct the amplitude basis that satisfies exchange symmetries, we need the (anti)symmetrization projector (i.e. Young operator), which is a function of the permutations. These permutations operate on both gauge and Lorentz structures. The representation matrices of the permutations can be expressed as the direct products of their respective representation matrices in the spaces of the gauge factor and Lorentz structures. Using these matrices, the (anti-)symmetrization projector is constructed. The resulting amplitude basis inherently satisfies the desired symmetry, yielding the final EFT operators. Detailed calculations are presented in Sec.~\ref{sec:exchange} and App.~\ref{sec:identical}. 




\section{Construction of the Basis}\label{sec:construction-basis}
\subsection{The structure of $\{\mathcal{C}\cdot F\}$}
The factorization in Eq.~(\ref{CFget}) indicates the amplitude basis involving massive particles can be classified into $\prod (2s_i + 1)$ configurations based on the polarization configurations $\{l_i\}$. Based on this classification,  
the massive amplitude basis can be constructed by first constructing the massless $c^{\{l_i\}}$ and $f^{\{l_i\}}$ basis to get the basis $\{c\cdot f\}^{\{l_i\}}$ and then massification. In the following, we will discuss the construction of $c^{{l_i}}$ and $f^{{l_i}}$ bases using Young diagrams.

\paragraph{$\mathcal{C}$ basis:}
$\mathcal{C}$ is a holomorphic function of the right-handed spinors, containing no left-handed spinors. As a result, it is inherently free from IBP redundancies caused by momentum conservation, as it contains no momentum factors. Additionally, since the little group indices of right-handed spinors from the same particle are fully symmetric, the square brackets formed by two such spinors vanish, i.e.~$[i^{\{I}i^{J\}}]=0$. Consequently,  $\mathcal{C}$ also has no EOM redundancies. The right-handed spinors from each polarization tensor belong to the $SU(2)_r$ representation with $(2s_i - l_i)$ fully symmetric \(\{\dot\alpha\}\) indices, corresponding to a $(2s_i - l_i + 1)$-dimensional representation.
$\mathcal{C}$ resides in the outer product of these $SU(2)_r$ representations, forming a reducible representation of $SU(2)_r$. 
Therefore, the irreducible representations decomposed from this outer product can be chosen as $\mathcal{C}$ basis, and we adopt the language of representation theory in Eq.~(\ref{Bpart}), where \(k\)-th represents the count of all Young tableaux from decomposition,
\begin{equation}
\mathcal{C}^{\{l_i\}} \subset \bigotimes_{\substack{i \text{-th} \\ \text{particle }}}
    \Yvcentermath1 \renewcommand\arraystretch{0.05} \setlength\arraycolsep{0.2pt}
    \underbrace{\yng(1)\ \cdots\ \yng(1)}_{2s_i-l_i} = \bigoplus_{\substack{k \text{-th} \\ \text{basis }}}\mathcal{C}^{\{l_i\}}_{k}.
    \label{Bpart}
\end{equation}
\paragraph{$F$ basis:}
For the $F^{\{l_i\}}$, it contains the left-handed spinors and momenta, so it must have the EOM and IBP redundancies. As discussed in~\cite{Dong:2022jru}, we can first construct its massless limit basis $f$ by semi-standard Young tableaux (SSYT) to remove both EOM and IBP redundancies.

Redundancies are removed for the following reasons: Since EOM can leads to a lower-dimensional $F$ basis with additional mass factors, \( \slashed{\bm{p}}_i | \bm{i}\rangle = m_i | \bm{i}] \), the massless limit \( F \to f \) removes terms containing mass factors. Furthermore, constructing \( f \) basis via the SSYT method can eliminate IBP redundancy. 

 Next, following the SSYT method in~\cite{Henning:2019enq,Dong:2021vxo}, we briefly review how to construct the complete set of $\{f\}$ bases without IBP redundancy. The massless spinors of $N$ external legs $\tilde{\lambda}_{\dot\alpha}^k \equiv \antiket{k}$ ($\lambda_{ k \alpha} \equiv  \ket{k}$) are embedded into the (anti-) fundamental representation of $U(N)$ symmetry with $k=1,\dots, N$.
So the basis of a $U(N)$ representation (i.e., a $U(N)$ SSYT) corresponds one-to-one to a polynomial of massless spinors. Conversely, this polynomial can also be systematically constructed through the SSYT according to the permutation symmetry of the $U(N)$ indices. For example, the scalar product of a right-/left-handed spinor pair can be obtained from \( U(N) \) SSYT with shape \( [1^2] / [1^{N-2}] \) (where \( [1^2] \) is shorthand for \( [1,1] \), and similarly \( [1^{N-2}] \) for \( [1,1,\dots,1] \)),
\begin{equation}
\begin{aligned}
\label{eq:YD_spinorproduct}
\Yvcentermath1 \ytableausetup{boxsize=1.3em, aligntableaux=center}
\begin{ytableau}
i \\ j \end{ytableau}
 \ &= \epsilon^{\dot{\alpha}\dot{\beta}} \tilde{\lambda}_{\dot{\alpha}}^i \tilde{\lambda}_{\dot{\beta}}^j = [ij], \\
\Yvcentermath1 N-2 \left\{\begin{array}{c} \textcolor{blue}{\begin{ytableau}
k_1 \\
k_2 \\
\none[\vdots] \\
\none[\vdots]
\end{ytableau}}
\end{array}\right.
&= \epsilon^{\alpha\beta} \lambda_{i,\alpha} \lambda_{j,\beta} \epsilon^{ijk_1 \ldots k_{N-2}} = \langle ij\rangle \epsilon^{ijk_1 \ldots k_{N-2}}\,.
\end{aligned}
\end{equation}
where $\varepsilon^{ij k_1 \ldots k_{N-2} }$ is the epsilon tensor. Note that the columns in the SSYT associated with the $U(N)$ indices of $\lambda$ are colored blue to distinguish them from the indices of $\tilde{\lambda}$.

The massless limit $f$ retains the same Lorentz representation as $F$ and must be in  the same $SU(2)_r$ representation with $\mathcal{C}$ in order to contract into Lorentz singlets. Thus, $f$ should be an $SU(2)_l$ singlet and an $SU(2)_r$ multiplet, with a spinor structure of the form $\langle\circ\circ\rangle^{L/2}[\circ\circ]^{r_2}|\circ]^{r_1-r_2}_{\dot\alpha}$. The independent $f$ basis with this form can be constructed by directly gluing the blue Young diagram (YD), with  $N-2$ rows and $L/2$ columns, and the white YD, with the shape of $[r_1,r_2]$, together without changing their shape, as shown in Eq.~(\ref{LR}). Other YDs decomposed from the out product of these two YDs, such as placing a white box below a blue one, are proportional to the total momentum $(\sum_{k=1}^N |k]\langle k|) = 0$ and therefore vanish.        
\begin{equation}
    \begin{aligned}
    f^{\{\dot\alpha\}}\left(\langle\circ\circ\rangle^{L/2}[\circ\circ]^{r_2}|\circ]^{r_1-r_2}_{\dot\alpha}\right)
    &=\Yvcentermath1 N-2 \left\{\rule{0pt}{3.6em} \begin{array}{c} \end{array} \right.\!\!\!\! \underbrace{  \blue{  \begin{array}{ccc}
    \yng(1,1)& \cdots & \yng(1,1) \\
    \vdots &  \ddots & \vdots \\
    \yng(1,1)& \cdots & \yng(1,1)
    \end{array}}  }_{L/2}  \!\otimes \,
    \renewcommand\arraystretch{0.05} \setlength\arraycolsep{0.2pt}\Yvcentermath1\begin{array}{c} \overbrace{\yng(1)\cdots\yng(1)\cdots\yng(1)}^{r_1}\\
    \underbrace{\yng(1)\cdots\yng(1)}_{r_2} {\color{white} \, \ \ \ \ \ \ \ } \end{array}\\
    &=\Yvcentermath1 N-2 \left\{\rule{0pt}{3.6em} \begin{array}{c} \end{array} \right.\!\!\!\!
     \begin{array}{ccccc }
\blue{ \yng(1,1) }& \blue \cdots &\blue{ \yng(1,1)} \yng(1,1)&  \cdots &\yng(1,1) \\
\blue\vdots &  \blue\ddots & \blue\vdots \quad \quad & & \\
\blue{ \yng(1,1)}& \blue\cdots  & \blue{ \yng(1,1)} {\color{white} \yng(1,1)}  & &
     \end{array}
     \!\!\!\!\!\!\!\!\!\!\!\!\!\!\!\!\!\!\!\!
     \!\!\!\!\!\!\!\!\!\!\!\!\!\!\!\!\!\!\!\!\!\!\!\!\!\!\!\!\!\!\!\!\!
     \renewcommand\arraystretch{0.05}\begin{array}{ccccccc}
\color{white}\yng(1) &\color{white}\cdots &\color{white}\yng(2) &\color{white}\cdots &\color{white}\yng(1) &\cdots  &\yng(1)\\
\color{white}\yng(1,1,1,1,1)&  &  &  &  &  &
     \end{array}
     \!\!\!\!\oplus  \cdots\,.
    \end{aligned}
    \label{LR}
\end{equation}
The complete $f$ basis then can be read off from the SSYTs of the first $U(N)$ YD. The number of fillings in each SSYT is determined by the helicities of the massless particles and the number of left-handed spinors, ${l_i}$, contained in the polarization tensors of the massive particles:
\begin{equation}
\begin{array}{rcll}
 \#\,i & = & \frac{L}{2} - l_i & \quad \text{for massive particle-$i$}, \\[8pt]
 \#\,j& = & \frac{L}{2} + 2h_j & \quad \text{for massless particle-$j$}.
\end{array}
\label{eq:filling_of_f}
\end{equation}

\subsection{Relationship of $\{\mathcal{C}\cdot F\}$ and $\{c \cdot f\}$}
\begin{figure}[H]
    \centering
    \begin{subfigure}{0.33\textwidth}
        \centering
        \includegraphics[width=\textwidth]{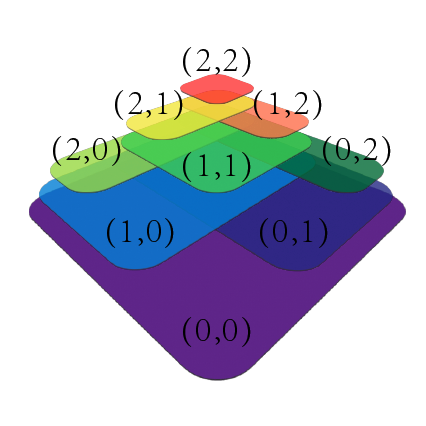}
        \caption{}
        \label{fig:mappingB}
    \end{subfigure}%
    \hspace{1.8cm}
    \begin{subfigure}{0.3\textwidth}
        \centering
        \includegraphics[width=\textwidth]{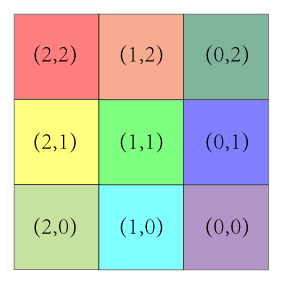}
        \caption{}
        \label{fig:mappingC}
    \end{subfigure}
    \caption{a) Subset relationships between $\{\mathcal{C} \cdot f\}^{\{l_1, l_2\}}$ for two massive vectors; 
    b) The top-down view of the left diagram, which corresponds to the lowest-dimensional amplitude basis $\{c \cdot f\}$. 
    The $\{\mathcal{C} \cdot f\}$ and the corresponding $\{c \cdot f\}$ bases are labeled by polarization configurations $(l_1,l_2)$  of two vectors. \( l_{1,2} = 0, 1, 2 \) represent right-handed transverse, longitudinal, and left-handed transverse configurations, respectively.}
    \label{fig:mapping}
\end{figure}

When contracting $\mathcal{C}$ and $f$ bases to form Lorentz singlets, no redundancies arise within each set of $\{\mathcal{C} \cdot f\}^{\{l_i\}}$. However, EOM redundancies may occur between different sets of $\{\mathcal{C} \cdot f\}^{\{l_i\}}$.
These arise when the right-handed spinors $\antiket{i^\prime}$ in $\mathcal{C}$ contract with the right-handed spinors $\antiket{i}$ from the momenta in $f$, producing a trivial overall mass factor.
As a result, some amplitudes in the set $\{\mathcal{C} \cdot f\}^{\{l_i\}}$ become equivalent to lower-dimensional ones in another set $\{\mathcal{C} \cdot f\}^{\{l'_i\}}$ through the following EOM, 
\begin{equation}
p_i^\mu \sigma_\mu |i'] = [ii'] |i\rangle \sim m_i |i\rangle\,.
\label{eq:Redun_between_Cf}
\end{equation}
Here, we use $i'$ to denote the spinors from the polarization in $\mathcal{C}$, to distinguish them from the spinors $i$ in $f$, so the factor $[ii']$ corresponds to the mass $m_i$ after performing massification $f\to F$. 
Sets with larger $l_i$ must include those with smaller $l_i^\prime$, as the EOM in Eq.~(\ref{eq:Redun_between_Cf}) causes number of right-handed spinors $l_i$ in $\mathcal{C}^{\{l_i\}}$ to decrease.
This establishes the subset relationships among different sets $\{\mathcal{C} \cdot f\}^{\{l_i\}}$, as shown in Fig.~\ref{fig:mappingB},
\begin{equation}
\{\mathcal{C}\cdot f\}^{\{l_i^\prime\}} \subset \{\mathcal{C}\cdot f\}^{\{l_i\}}\,,\quad l_i\leq\, l_i^\prime\,,
\label{CFsubset}
\end{equation}
which can be rigorously proven using representation theory and linear algebra by analyzing the ranks of linear spaces. With the invariant rank of the series of linear spaces, which start with the maximal space $\{\mathcal{C} \cdot f\}^{\{l=0\}}$ and successively add direct sums of spaces $\{\mathcal{C} \cdot f\}^{\{l_i\}}, l_i = 1,2,\dots$, the relationships among subspaces will be obvious.

Fig.~\ref{fig:mappingB} illustrates the relations between the basis sets with different configurations $\{l_i\}$ for the case of two massive vectors. The blocks with different colors represent distinct $\{\mathcal{C} \cdot f\}^{\{l_i\}}$ sets. The number $(i,j)$ represents the polarization configuration $(l_1,l_2)$ of two vectors. 
And the overlapping area between the two blocks represents their amplitude bases equivalent by EOM. 
So the biggest purple block, $\{\mathcal{C} \cdot f\}^{\{0,0\}}$, is the complete and independent EFT operator basis, though some of its operators can be converted into the lower-dimensional ones through EOM. 
The subset relation in Eq.~(\ref{CFsubset}) ensures that when operators in different sets $\{\mathcal{C} \cdot f\}^{\{l_i\}}$ are EOM equivalent, those in the upper layer of Fig.~\ref{fig:mappingB} always have lower operator dimensions. 
Thus, looking from top to bottom in Fig.~\ref{fig:mappingB}, all unobstructed blocks are the lowest-dimensional complete basis, as shown in Fig.~\ref{fig:mappingC}.
All the bases in the obstructed blocks contain mass factor like \( [ii'] \). To eliminate them, it is sufficient to replace all \( i' \) in \( \mathcal{C} \) with \( i \),  
\begin{equation}
\{\mathcal{C} \cdot f\}^{\{l_i\}} \xrightarrow{i' \to i} \{c \cdot f\}^{\{l_i\}}\,,
\end{equation}
which is equal to taking the massless limit of \( \mathcal{C} \), denoted as \( c \). And this gives us the lowest-dimensional basis for the massive particles \footnote{Alternative approaches, such as~\cite{Song:2023jqm,Song:2023lxf}, construct longitudinal configurations using massless scalar bases satisfying the Adler zero condition. However, these methods are restricted to massive vectors and require additional enumeration to identify the basis.}. 

\subsection{Young tableau for construction $\{c\cdot f\}$}
In this subsection, we will discuss how to efficiently construct the massless basis $\{c \cdot f\}^{\{l_i\}}$. 
The traditional method to construct $\{c \cdot f\}^{\{l_i\}}$, as shown in Fig.~\ref{fig:contract}, is to first construct the massless $c$ and $f$ basis separately and then combine them into Lorentz singlets. Since the $\{c \cdot f\}^{\{l_i\}}$ basis can be obtained by taking the massless limit of $\{\mathcal{C} \cdot f\}^{\{l_i\}}$ basis, and the massless limit of $\{\mathcal{C} \cdot f\}^{\{l_i\}}$ is equivalent to stripping the LG index of the massive spinors in $\mathcal{C}$, thus, we can first construct $\{\mathcal{C} \cdot f\}^{\{l_i\}}$  basis and then derive the $\{c \cdot f\}^{\{l_i\}}$ basis from $\{\mathcal{C} \cdot f\}^{\{l_i\}}$. 
 
We find that the SSYTs of the basis $(\mathcal{C} \cdot f)$ can be directly constructed if we use the box labeled by number $i'$ in the SSYT to represent the spinor $\antiket{i^I}$ in $\mathcal{C}$. In this setup, the numbers in the SSYT of $\{\mathcal{C} \cdot f\}^{\{l_i\}}$, ordered as $1 < \dots < N < N' < \dots < 1'$, naturally distinguish the $\mathcal{C}$ and $f$'s tableaux. If we first construct this kind of SSYTs, $\{\mathcal{C} \cdot f\}^{\{l_i\}}$ basis can be directly read off. Then taking the massless limit $\mathcal{C} \to c$ directly yields the desired basis $\{c \cdot f\}^{\{l_i\}}$. Based on these discussions, the SSYTs of the amplitude basis $\{c \cdot f\}^{\{l_i\}}_d$ with operator dimension-$d$ can be constructed as follows:
\begin{figure}[htb]
    \centering
    \includegraphics[width=0.6\linewidth]{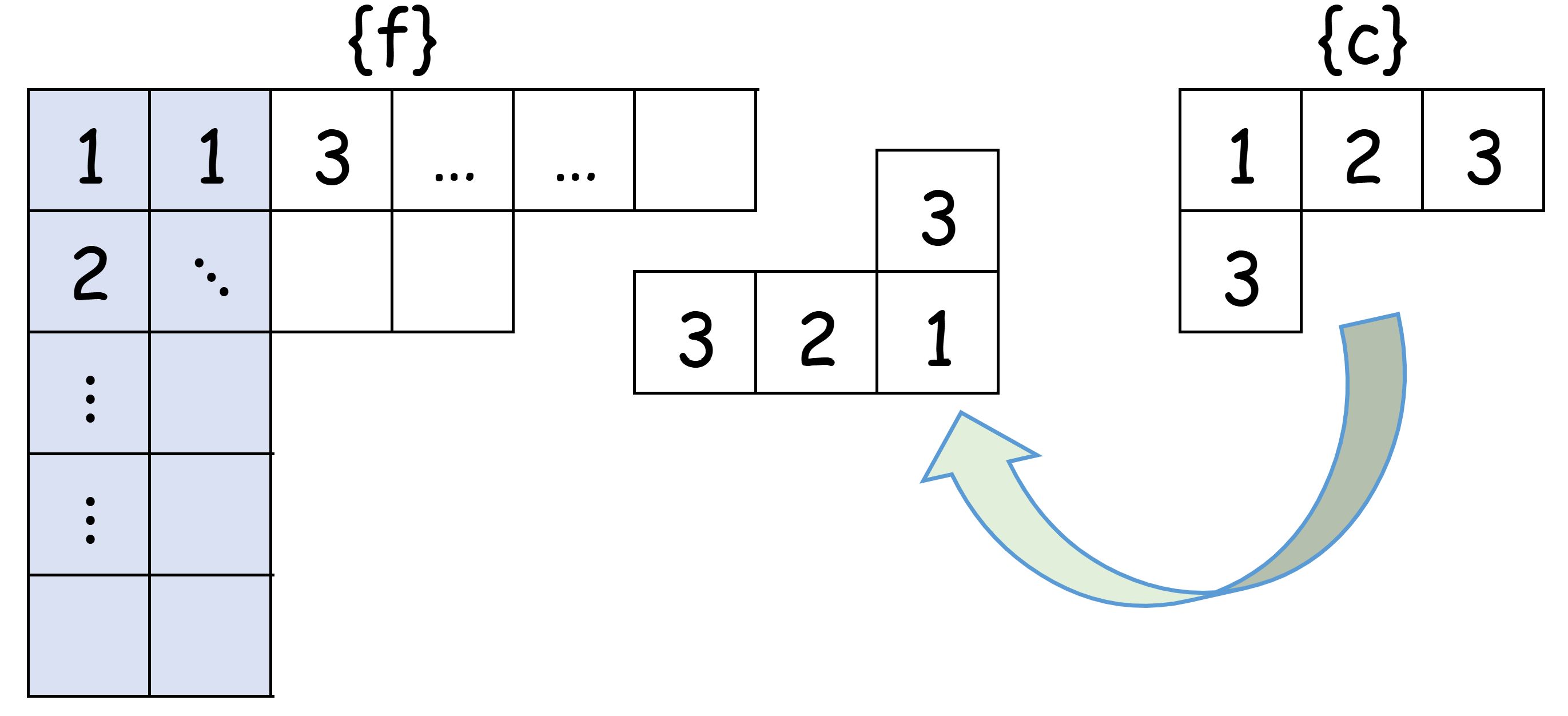}
    \caption{Rotate the SSYT of \( c \) by $180^\circ$ and attach it to \( f \). The corresponding amplitude basis is obtained by contracting the \( \dot\alpha \) indices of the spinors in \( c \) and \( f \) within the same column.}
    \label{fig:contract}
\end{figure}
\begin{itemize}
    \item[$\bullet$] \textit{Young diagram}: First, we construct the shape of the SSYT, i.e. its Young diagram. According to Eq.~(\ref{eq:YD_spinorproduct}), the total numbers of left-handed spinors \(L\) and right-handed spinors \(R\) determine the number of blue and white columns in the Young diagram, respectively. The total numbers are given by:
    \begin{equation}
    \begin{aligned}
    \label{eq:YCF}
    R &= d-N+n_A + \sum (s_i-l_i) + \sum h_j, \\
    L &= d-N+n_A - \sum (s_i-l_i) - \sum h_j,
    \end{aligned}
    \end{equation}
    where $n_A$ is the number of bare fields, $A_\mu \sigma^\mu$ or $\psi_\mu^{L/R} \sigma^\mu$ in Table~\ref{tab:dm_buildingblocks}. The corresponding Young diagram is $[(L+R)/2, (L+R)/2, (L/2)^{N-4}]$.

    \item[$\bullet$] \textit{Filling numbers}: Next, using Eqs.~(\ref{eq:filling_of_f}) and (\ref{Bpart}), we determine the number of fillings for massive particle-$i$ and massless particle-$j$ in the Young diagram as:
    \begin{equation}
    \begin{aligned}
    & \#\,i = \frac{L}{2} - l_i, \quad \#\,i' = 2s_i-l_i, \\
    & \#\,j = \frac{L}{2} + 2h_j,
    \end{aligned} \label{eq:filling}
    \end{equation}
    where $i'$ is restricted to the $R/2$ white columns since it originates from $\mathcal{C}$. The numbers are filled in the order $1 < \dots < N < N' < \dots < 1'$.

    \item[$\bullet$] \textit{Massless limit of polarizations}: Finally, replace all $i'$ with $i$ and retain only the linearly independent $\{c \cdot f\}$ to form the basis. This step is implemented by semi-standardizing the Young tableaux after the $i' \to i$ replacement and removing any zero or linearly dependent terms.
\end{itemize}

\subsection{Exchange symmetries on $\{c\cdot f\}$} \label{sec:exchange}
In this subsection, we explain a more efficient method to construct the amplitude bases involving identical particles, compared to the previous method~\cite{Dong:2021vxo,Dong:2022jru,Dong:2022mcv}. 
The exchange symmetry of identical particles swaps their momenta \(\bm{p}_i\) and polarizations \(\bm{\varepsilon}^{l_i}_i\). For the amplitude basis, this corresponds to exchanging the associated spinors and LG indices while preserving the operator dimension \(d\).

The amplitude basis involving two identical dark photons is used as an example. 
As discussed above, the representation matrix of exchange symmetry between the two photons, denoted as $\hat{P}_{(12)}$, can be obtained by acting it on the $\{c \cdot f\}^{\{l_1, l_2\}}_d$ basis,
\begin{equation}
    \hat{P}_{(12)} \begin{pmatrix}
        \{c \cdot f\}^{\{l_1, l_2\}}_d \\
        \{c \cdot f\}^{\{l_2, l_1\}}_d
    \end{pmatrix} =
    \begin{pmatrix}
        \{c \cdot f\}^{\{l_2, l_1\}}_d \\
        \{c \cdot f\}^{\{l_1, l_2\}}_d
    \end{pmatrix} \equiv M_{(12)} \begin{pmatrix}
        \{c \cdot f\}^{\{l_1, l_2\}}_d \\
        \{c \cdot f\}^{\{l_2, l_1\}}_d
    \end{pmatrix} = \begin{pmatrix}
        0 & 1\\
        1 & 0
    \end{pmatrix}\begin{pmatrix}
        \{c \cdot f\}^{\{l_1, l_2\}}_d \\
        \{c \cdot f\}^{\{l_2, l_1\}}_d
    \end{pmatrix}.
\end{equation}
Therefore, the matrix of the Young operator $\mathcal{Y}_{\text{\tiny\young(12)}} = (\mathds{1} + \hat{P}_{(12)})/2$ of the exchange symmetry $(12)$ is given by $\mathcal{Y}_{\text{\tiny\young(12)}}=(\mathds{1} + M_{(12)})/2$ as
\begin{equation}
\label{equ:identical-projection-noneq}
    \mathcal{Y}_{\text{\tiny \young(12)}} \begin{pmatrix}
        \{c \cdot f\}^{\{l_1, l_2\}}_d \\
        \{c \cdot f\}^{\{l_2, l_1\}}_d
    \end{pmatrix} \ = \frac{1}{2}\begin{pmatrix}
        1 & 1\\
        1 & 1
    \end{pmatrix}\begin{pmatrix}
        \{c \cdot f\}^{\{l_1, l_2\}}_d \\
        \{c \cdot f\}^{\{l_2, l_1\}}_d
    \end{pmatrix} 
    = \frac{1}{2}\begin{pmatrix}
        \{c \cdot f\}^{\{l_1, l_2\}}_d + \{c \cdot f\}^{\{l_2, l_1\}}_d \\
        \{c \cdot f\}^{\{l_1, l_2\}}_d + \{c \cdot f\}^{\{l_2, l_1\}}_d
    \end{pmatrix}\,.
\end{equation}
Thus, constructing either $\{c \cdot f\}^{\{l_2, l_1\}}_d$ or $\{c \cdot f\}^{\{l_1, l_2\}}_d$ yields the same result after symmetrization. So it suffices to construct only one of them when \( l_1 \neq l_2 \).

However, when \( l_1 = l_2 \), we need to carefully calculate the representation matrix of Young operator $\mathcal{Y}$ on the set \(\{c \cdot f\}^{\{l, l\}}_d\). Since \(\{c \cdot f\}^{\{l, l\}}_d\) is closed under identical permutation, the \(\{c \cdot f\}^{\{l, l\}}_d\) bases and the resulting bases after imposing the $\hat{P}_{(12)}$ operation can both be decomposed to the same set of the amplitude bases from SSYTs via semi-standardization
\begin{equation}
\{c \cdot f\}^{\{l_i\}}_d \xrightarrow{\text{semi-standardization}} \{e^{SSYT}\} \xleftarrow{\text{semi-standardization}}  \hat{P}_{(12)}\{c \cdot f\}^{\{l_i\}}_d.
\end{equation}
The representation matrix \(M_{(12)}\) is determined by the coefficient matrices $A,B$ of the SSYT decomposition before and after the permutation, through the linear relation:
\begin{equation}
(c\cdot f)_i = A_{ij} e_j^{SSYT}, \quad \hat{P}_{(12)}(c\cdot f)_i = B_{ij} e_j^{SSYT}, \quad B_{ij} = M_{(12),\,ik} A_{kj}\,,
\end{equation}
where $e_j^{SSYT}$ is the massless amplitude basis from SSYT. Note that a $(c\cdot f)$ basis does not generally correspond to a SSYT of the $U(N)$ group.    

For the example in Fig.~\ref{fig:mappingC}, the exchange symmetry between identical particles 1 and 2 eliminates half of the blocks with unequal polarization configurations \((l_1, l_2)\), retaining only those with \(l_1 \geq l_2\) without loss of generality. The matrix representation of Young operator of permutation symmetry $(12)$, $Y_{\text{\tiny\young(12)}} = (\mathds{1} + M_{(12)})/2$, applies only on diagonal blocks where \(l_1 = l_2\). As a result, the nine initial blocks are reduced to six: three non-diagonal blocks that survive after symmetrization, and three diagonal blocks that require additional projection calculations using \(\mathcal{Y}_{\text{\tiny\young(12)}}\) to symmetrize. The case for identical particles with gauge factors is discussed in App.~\ref{sec:identical}.


\section{Example: $XX e_R e_R^\dagger$ Amplitude and Operator Basis}\label{sec:example}
In this section, we explicitly construct the operator basis at dimension $d=8$ for $XXe_R e_R^\dagger$ to demonstrate above method. First, we determine the polarization configurations of $\{\mathcal{C}\cdot F\}^{\{l_i\}}$ . Since the massive vector has three polarizations, $3\times3$ configurations with $l_{1,2} = 0,1,2$ are involved in this case. 
To distinguish the bases at different amplitude dimensions, the basis set can be classified by the dimension of its amplitude basis $d_a$ and the polarization configuration $l \equiv (l_1,l_2)$.     
Therefore, we have basis sets labeled by $\{d_a, l\}$, such as $\{d_a, l\} = \{5, (0,1)\}$ and $\{d_a, l\} = \{6, (1,1)\}$.

Before constructing the amplitude basis in each set, we utilize the exchange symmetry of identical particles to project out some independent sets. For example, we can construct the basis with $l = (0,2)$ to represent both $l = (0,2)$ and $l = (2,0)$ since these particles are identical, as explained in Sec.~\ref{sec:exchange}. After this filtering, there remain six independent sets, as follows:
\begin{equation}
       \begin{aligned}
        &x_1 = \{4,(0,0)\}\,,\,
        x_2 = \{4,(0,2)\}\,,\,
        x_3 = \{4,(2,2)\}\,, \,\\
        &x_4 = \{5,(0,1)\}\,,\,
        x_5 = \{5, (1,2)\}\,,\,
        x_6 =\{6, (1,1)\}\,.
       \end{aligned}
\end{equation}
Here the first element in the $x_i$ means the amp dimension, the difference between amplitude dimension and operator dimension comes from the EOM of massive particles.
For the sets with $l_1 =l_2$, as mentioned in Sec.~\ref{sec:exchange}, the symmetry of identical particles further imposes restrictions on the sets of $x_{1,3,6}$, requiring the bases to be the eigenstates of the Young operator $\mathcal{Y}_{\text{\tiny\young(12)}}$. 

Using the method outlined in the previous section, we can derive all $d=8$ EFT operators for $XXe_R e_R^\dagger$ with a specific polarization configuration $x_i$, denoted as $\mathcal{L}^{x_i}_{\pd=8} (XXe_R e_R^\dagger)$. We can first construct the SSYTs and then obtain $\{\mathcal{C} \cdot f\}^{x_i}$ bases, following the rules in Eq.~(\ref{eq:YCF}) and (\ref{eq:filling}). Then take the massless limit of $\{\mathcal{C} \cdot f\}^{x_i}$ bases, choose the independent massless bases, and finally the independent massive amplitude bases without the exchange symmetry of identical particles  can be obtained by massification,

\begin{equation}
    \label{eq:XXeebarbasis}
    \begin{aligned}
    \renewcommand{\arraystretch}{1.2}
        &
        \{ \mathcal{L} \}^{x_1}_{\pd=8}=\!
        \begin{array}{l}
             \sbk{\mOne \mTwo} \sbk{\mOne 3}  \sbk{\mTwo 3} \abk{34}
        \end{array}
        \,,
        &
        \{ \mathcal{L} \}^{x_2}_{\pd=8}=\!
        \begin{array}{l}
            \sbk{\mOne 3} \sbk{\mOne 4} \abk{\mTwo 4}^2
        \end{array}
        \\
        &
        \{\mathcal{L} \}^{x_3}_{\pd=8}=\!
            \begin{array}{l}
                \abk{\mOne \mTwo}\abk{\mOne 4}\abk{\mTwo 4}\sbk{34}
            \end{array}
            \,,
        &
        \{\mathcal{L} \}^{x_4}_{\pd=8}=\!\left \{
        \begin{array}{l}
            \sbk{\mOne 3}^2\sbk{3\mTwo}\abk{\mTwo 3}\abk{34}\\
            \abk{\mTwo 3}\abk{34}\sbk{34}\sbk{\mOne 3}\sbk{\mOne \mTwo}
        \end{array}
        \right \},\\
        &
        \{\mathcal{L} \}^{x_5}_{\pd=8}=\!\left \{
            \begin{array}{l}
                \abk{\mOne \mTwo}\abk{\mTwo 4}\abk{34}\sbk{34}\sbk{\mOne  3}\\
                \abk{\mOne 4}\abk{\mTwo 4}^2\sbk{34}\sbk{\mOne 4} 
            \end{array}
            \right \}\,,
        &
        \{\mathcal{L} \}^{x_6}_{\pd=8}=\!\left \{
        \begin{array}{l}
            \abk{\mOne \mTwo}\abk{34}^2\sbk{34}\sbk{\mOne 3}\sbk{\mTwo3}\\
            \abk{\mOne 4}\abk{\mTwo 4}^2\sbk{\mTwo 3}\sbk{\mOne 4}\sbk{\mTwo4}\\
            \abk{\mOne 4}\abk{\mTwo 4}^2\sbk{\mTwo 4}\sbk{34}\sbk{\mOne \mTwo}
        \end{array}
        \right \}.
    \end{aligned}
\end{equation} 

Next, use the above independent bases to construct the bases satisfying the exchange symmetry of two dark photons.   
We can first apply the exchange operator $\hat{P}_{(12)}$ to the basis $x_1$ and $x_3$, and get
\begin{equation}
   \hat{P}_{(12)} \{\mathcal{L} \}^{x_{1,3}}_{\pd=8} = - \{\mathcal{L} \}^{x_{1,3}}_{\pd=8}.
\end{equation}
Therefore, we have the matrix representation of Young operator as $\mathcal{Y}_{\text{\tiny\young(12)}}^{x_{1,3}} = (\mathds{1} + M^{x_{1,3}}_{(12)})/2 = 0$, which means $\{\mathcal{L} \}^{x_{1,3}}$ does not satisfy the exchange symmetry and thus does not contribute to the EFT of $XXe_R e_R^\dagger$.

Similarly, the permutation $\hat{P}_{(12)}$ acting on the basis $x_6$ results in
\begin{equation}
    M_{(12)}^{x_6} =
    \renewcommand{\arraystretch}{1.2}
    \setlength{\arraycolsep}{3pt}
    \begin{pmatrix}
        -1 & 0 & 0 \\
        0 & -1 & 0 \\
        1 & 0 & 1
    \end{pmatrix},
    \quad
    Y_{\text{\tiny\young(12)}}^{x_6} =\frac{1}{2} \qty(\mathds{1}+  M_{(12)}^{x_6}) =
    \begin{pmatrix}
        0 & 0 & 0 \\
        0 & 0 & 0 \\
        \frac{1}{2} & 0 & 1
    \end{pmatrix},
\end{equation}
where $\mathcal{Y}_{\text{\tiny\young(12)}}^{x_{6}}$ is the representation matrix of Young operator $\mathcal{Y}_{\text{\tiny\young(12)}}$ in the space of basis set $x_6$. 
Since the rank of the representation matrix $\mathcal{Y}_{\text{\tiny\young(12)}}^{x_{6}}$ is 1, there is only one amplitude in $\mathcal{L}^{x_6}$ that satisfies the exchange symmetry. By calculating the eigenstates with nonzero eigenvalues of $\mathcal{Y}_{\text{\tiny\young(12)}}^{x_6}$, we find that this amplitude can be expressed as $\mathcal{Y}_{\text{\tiny\young(12)}} \abk{\mOne 4}\abk{\mTwo 4}^2\sbk{\mTwo 4}\sbk{34}\sbk{\mOne \mTwo}$. 

The exchange symmetry can project on bases labeled by $x_2=\{4,(0,2)\}$ and bases labeled by $x_2^\prime = \{4,(2,0)\}$. Because of the fact $\mathcal{Y}_{\text{\tiny\young(12)}} \{\mathcal{L}\}^{x_2} = \mathcal{Y}_{\text{\tiny\young(12)}} \{\mathcal{L}\}^{x_2^\prime} $ as discussed in Eq.~(\ref{equ:identical-projection-noneq}), $\mathcal{Y}_{\text{\tiny\young(12)}} \{\mathcal{L}^{x_2}\}$ can represent the contribution from $x_2$ and $x_2^\prime$ under the exchange symmetry of particles 1 and 2 and the process is similar for $x_4$ and $x_5$ since they also have $l_1 \neq l_2$.

We have discussed the projection on each $x_i$ basis. We have three non-diagonal bases as $\mathcal{Y}_{\text{\tiny\young(12)}}\{\mathcal{L}\}^{x_2}$, 
$\mathcal{Y}_{\text{\tiny\young(12)}}\{\mathcal{L}\}^{x_4}$, 
$\mathcal{Y}_{\text{\tiny\young(12)}}\{\mathcal{L}\}^{x_5}$ contribute to the independent EFT basis under exchange symmetry, and one diagonal bases $\mathcal{Y}_{\text{\tiny\young(12)}} \{\mathcal{L}\}^{x_6}$ also contributes, which is $\mathcal{Y}_{\text{\tiny\young(12)}} \abk{\mOne 4}\abk{\mTwo 4}^2\sbk{\mTwo 4}\sbk{34}\sbk{\mOne \mTwo}$. 
Finally, the complete basis at $d =8$ is the collection of all the above $\mathcal{Y}^{x_i}_{\text{\tiny\young(12)}}\{\mathcal{L}^{x_i}\}$,
\begin{equation}
    \mathcal{Y}_{\text{\tiny\young(12)}}\{\mathcal{L}\} = \mathcal{Y}_{\text{\tiny\young(12)}}\{\mathcal{L}^{x_2},\mathcal{L}^{x_4},\mathcal{L}^{x_5},\mathcal{L}^{x_6}\} 
        = \mathcal{Y}_{\text{\tiny\young(12)}} \!\left \{
        \begin{array}{l}
            \sbk{\mOne 3} \sbk{\mOne 4} \abk{\mTwo 4}^2\\
            \sbk{\mOne 3}^2\sbk{3\mTwo}\abk{\mTwo 3}\abk{34}\\
            \abk{\mTwo 3}\abk{34}\sbk{34}\sbk{\mOne 3}\sbk{\mOne \mTwo}\\
            \abk{\mOne \mTwo}\abk{\mTwo 4}\abk{34}\sbk{34}\sbk{\mOne  3}\\
            \abk{\mOne 4}\abk{\mTwo 4}^2\sbk{34}\sbk{\mOne 4} \\
            \abk{\mOne 4}\abk{\mTwo 4}^2\sbk{\mTwo 4}\sbk{34}\sbk{\mOne \mTwo}
        \end{array}
        \right \}. \label{eq:DPbasis}
\end{equation} 
The Young operator means that bases are symmetrized under the exchange symmetry between dark photons $1$ and $2$.

Finally, we can map this amplitude basis to the EFT operators. For instance, based on the correspondences in Tables~\ref{tab:sm_buildingblocks} and \ref{tab:dm_buildingblocks}, the operator corresponding to the first amplitude basis in Eq.~(\ref{eq:DPbasis}) above is given by 
\begin{equation}
    \mathcal{Y}_{\text{\tiny\young(12)}}\sbk{\mOne 3} \sbk{\mOne 4} \abk{\mTwo 4}^2 \rightarrow 
    \left( \left( {e}_{R}^{} \right) \sigma^{\mu\nu} \bar{\sigma}^{\rho} \sigma^{\xi\tau} \left( D_{\rho} {e}_{R}^{\dagger} \right) \right) {X}_{\mu \nu}^{+} {X}_{\xi \tau}^{-}
\end{equation}

In summary, this method enables the construction of EFT operators for particles with arbitrary mass and spin. These operators have been incorporated into a Mathematica package, which we utilized to systematically generate EFT operators for the interactions between the dark photon and SM particles, as well as those between the gravitino like and SM particles. The resulting operators are listed in tabular form in Appendices~\ref{app:DP} and \ref{app:Gravitino}, respectively.
\section{Conclusion}\label{sec:conclusion}

This paper presents a systematic, efficient, and redundancy-free method for constructing EFT operator bases, applicable to theories involving massive and high-spin particles for effective operator construction and phenomenological studies. Our construction is based on the on-shell amplitude method and the Young tensor technique, which leverages the connection between low-energy EFT and the high-energy limit to reduce complexity and avoid introducing extra Goldstone modes, as did in~\cite{Song:2023jqm,Song:2023lxf}. Compared to the previous methods, our method is more efficient. Moreover, it is applicable to EFTs involving particles with spin $>1$ and provides a tool for exploring EFT of new physics models.

In this work, we construct the complete EFT operator basis for the dark photon and also for spin $3/2$ particles up to dimension 8, as shown in App.~\ref{app:DP} and \ref{app:Gravitino}. The corresponding \textit{Mathematica} package for operator generation has been made publicly available on GitHub (\url{https://github.com/zizhengzhou/MassiveAmplitude/tree/dark-matter}), enabling users to generate EFT operators for their models of interest. These operators serve as powerful tools for phenomenological studies and consistent EFT calculations by offering a complete effective action.


Additionally, the operators at the higher dimension are crucial for exploring experimental signals beyond leading-order operators. In certain UV scenarios, higher-dimensional operators may produce signals comparable to those generated by leading-order operators, especially when the latter are suppressed due to symmetries or accidental cancellations. 
This underscores the importance of a complete EFT operator basis, which enables comprehensive exploration of the Wilson coefficient parameter space using S-matrix bootstrap techniques, collider experiments, and astronomical observations. 
Based on power counting rules, the effects of next-to-leading-order operators are expected to be comparable to the one-loop corrections of leading-order operators, a topic that has garnered considerable attention in recent years.

Our method provides a foundation for extending EFT operator constructions to more complex particle models. 
These advancements bridge the gap between theoretical models and experimental searches,  contributing to new physics discoveries in dark matter and beyond.

\acknowledgments
T.M. is partly supported by the Yan-Gui Talent Introduction Program (grant No. 118900M128) and the Chinese Academy of Sciences Pioneer Initiative ``Talent Introduction Plan". Z.D. thanks Alex Pomarol for his useful advice and is partly supported by the research grants 2021-SGR-00649 and PID2023-146686NB-C31, funded by MICIU/AEI/10.13039/501100011033/ and FEDER, UE.

\appendix
\appendix
\section{Conventions}
\label{sec:convention}
We use the following conventions for metric and sigma algebra in the expressions of EFT basis. 
\begin{equation}
    \begin{aligned}
        &g_{\mu\nu} = \operatorname{diag}(+1, -1, -1, -1) \\
        &\epsilon^{0123}=-\epsilon_{0123}=+1\,,\quad \epsilon^{12}=\epsilon_{21}=+1 \,,\\
        &\sigma^{\mu}_{\alpha\dot{\alpha}} = (1, \vec{\sigma})\,,\quad \bar\sigma^{\mu\ \dot{\alpha}\alpha }=\epsilon^{\alpha\beta}\epsilon^{\dot{\alpha}\dot{\beta}}\sigma^{\mu}_{\beta\dot{\beta}}= (1, -\vec{\sigma}) \,,\\ 
        &\sigma^\mu_{\alpha\dot{\alpha}} \bar\sigma^{\nu \dot{\alpha}\beta} + \sigma^\nu_{\alpha\dot{\alpha}} \bar\sigma^{\mu \dot{\alpha}\beta} = 2 g^{\mu\nu} \delta_{\alpha\beta}\\
        &\Tr\qty(\sigma^\mu \bar \sigma^\nu) = 2 g^{\mu\nu} \,,\\
        &\sigma^{\mu\nu}_{\alpha\beta} = \frac{i}{2}\qty(\sigma^\mu_{\alpha\dot{\alpha}} \bar\sigma^{\nu\,\dot{\alpha}}_{\ \ \ \beta} - \sigma^\nu_{\alpha\dot{\alpha}} \bar\sigma^\mu_{\dot{\alpha}\beta}) \,,\\
        &\bar\sigma^{\mu\nu\ \dot{\alpha}\dot{\beta}} = \frac{i}{2}\qty( \bar\sigma^{\mu\ \dot{\alpha}\alpha} \sigma^{\nu\ \dot{\beta}}_{\alpha} - \bar\sigma^{\nu\ \dot{\alpha}\alpha} \sigma^{\mu\ \dot{\beta}}_{\alpha}).
    \end{aligned}
\end{equation}
Using the above definitions and amplitude-operator correspondence, for massive particles, we need spinors with symmetrized free little group indices as
\begin{equation}
    \begin{aligned}
        &p_\mu \sigma^\mu_{\alpha\dot\alpha} = 
        \ket{i_I}_\alpha [i^I|_{\dot \alpha}\,,
        \quad p_\mu \bar\sigma^{\mu\ \dot\alpha\alpha} = 
        |i^I]^{\dot\alpha}\bra{i_I}^\alpha\,,\\
        &\psi_{L,\alpha} = \ket{i_I}_\alpha\,,\ \psi_{R,\alpha} = 
        |i_I]^{\dot\alpha}\,, \\
        &F_{\mu\nu}^{+}\bar{\sigma}^{\mu\nu,\dot{\alpha} \dot{\beta}} = 
        |i_{\{ I }]^{\dot\alpha}|i_{ I \} }]^{\dot\beta}\,,
        \quad  F_{\mu\nu}^{-}{\sigma}^{\mu\nu}_{\alpha\beta} = 
        \ket{i_{\{ I_1 }}_{\alpha}\ket{i_{ I_2 \} }}_{\beta}\,, \\
        &m A_\mu \sigma^\mu_{\alpha\dot\alpha} = 
        \ket{i_{\{I_1}}_\alpha [i_{I_2\}}|_{\dot \alpha}\,,
        \quad m A_\mu \bar\sigma^{\mu,\dot\alpha\alpha} = 
        |i_{\{I_1}]^{\dot\alpha}\bra{i_{I_2\}}}^\alpha\,,\\
        &\partial_\mu \psi_{L\,\alpha,\nu}\sigma_{\alpha\beta}^{\mu\nu} = 
        \ket{i_{\{I_1}}_\alpha \ket{i_{I_2}}_\beta  \ket{i_{I_3\}}}_\gamma\,, \\
        &m \psi_{L\,\alpha, \mu}\sigma_{\beta}^{\mu\,\dot{\beta}} = 
        \ket{i_{\{I_1}}_\alpha \ket{i_{I_2}}_\beta  |i_{I_3\}}]_{\dot{\beta}}.
    \end{aligned}
\end{equation}
We also have right-hand spinors for massive $3/2$ particles. Wave functions as $\psi_L$ should be considered as $\psi_R^\dagger$ and vice versa when it represents an anti-fermion as an ingoing particle since one can use the crossing symmetry to flip it. For massless particles LG indices on spinors are removed.

Weyl spinor definitions as $\psi_L = P_L\Psi, \psi_R = P_R\Psi$ (0 components removed) and the charge conjugation matrix $C$ allow spinors to be translated into 4-spinor forms as 
\begin{equation}
    \begin{aligned}
        \text{2-spinor} & \quad \text{Fermion} & \quad \text{Anti-fermion} \\
        \ket{i}        & \quad P_L \Psi        & \quad CP_R \overline{\Psi}^T \\
        \bra{i}        & \quad \Psi^T P_L C \Psi & \quad \overline{\Psi}_R \\
        |i]            & \quad P_R \Psi        & \quad CP_L \overline{\Psi}^T \\
        [i|            & \quad \Psi^T P_R C \Psi & \quad \overline{\Psi}_L \\
    \end{aligned}
\end{equation}

We also have 
\begin{equation}
    \begin{aligned}
        &C=i\gamma^0\gamma^2=\begin{pmatrix} \epsilon_{\alpha\beta}&0\\0&\epsilon^{\dot{\alpha}\dot{\beta}}\end{pmatrix}\,,\\
        &\gamma^{\mu}=\begin{pmatrix}
            0&\sigma^{\mu}_{\alpha\dot{\beta}}\\\bar{\sigma}^{\mu\dot{\alpha}\beta}&0
        \end{pmatrix}\,,\\
        &\sigma^{\mu\nu}=\dfrac{i}{2}[\gamma^\mu,\gamma^\nu]=\begin{pmatrix}
            \left(\sigma^{\mu\nu}\right)_\alpha{}^\beta&0\\0&\left(\bar{\sigma}^{\mu\nu}\right)^{\dot{\alpha}}{}_{\dot{\beta}}
        \end{pmatrix}
        \end{aligned}
\end{equation}

\section{Identical Particles with Gauge Factor}
\label{sec:identical}
Symmetries of identical particles act as projections onto the bases. For \(n\) identical bosons or fermions in the same polarization configuration, the exchange symmetry constrains their representations under the permutation group \(S_n\) as,
\begin{equation}
[r_n] = [n] \quad \textit{for bosons}\,, \quad\quad\quad\quad [r_n] = [1^n] \quad \textit{for fermions}\,.
\end{equation}
Applying the symmetrization Young operator $\mathcal{Y}_{[r_n]}$ to the complete basis set $\{\mathcal{M}\}$ yields the representation matrix $M_{[r_n]}$ and projects out the operators satisfying the exchange symmetry. Note that any Young operator of \(S_n\) can be expressed as a function of the permutation elements \((12)\) and \((1\ldots n)\), so it is sufficient to compute the representation matrices for these two.

The amplitude basis is composed of a gauge factor and the Lorentz structure,
\begin{equation}
\{\mathcal{M}\} \xrightleftharpoons[\rm Massification]{\rm Massless\; limit} \{T\}\times \{c\cdot f\}\,.
\end{equation}
Therefore, it suffices to compute the representation matrices $M^T_{(12),(1,\ldots,n)}$ and $M^{cf}_{(12),(1,\ldots,n)}$ in the $\{T\}$ and $\{c \cdot f\}$ space respectively, for the required Young operator of $S_n$. In conclusion, the representation matrix $M_{[r_n]}$ is determined by computing the Kronecker product of the matrices $M^T$ and $M^{cf}$. The (anti)symmetric representation matrix is
\begin{equation}
 \label{eq:B/Fmatrix}
 \begin{aligned}
    &M_{[r_n]}=\mathcal{Y}_{[r_n]}(x,y)\,,\\
    &x=M^T_{(12)}\otimes M^{cf}_{(12)}\,,\quad y=M^T_{(1\ldots n)}\otimes M^{cf}_{(1\ldots n)}\,.
    \end{aligned}
\end{equation}
$\{T\}$ here is the extra gauge structures. As the computation of $M^{cf}$ has already been introduced in the main text, we briefly review the construction of independent gauge bases and the process of obtaining the representation matrices on $\{T\}$. This is demonstrated through the systematic construction of $SU(2)_L$ gauge factors $T$ and the $M^{T}$ matrix, which is also applicable to general $SU(N)$ gauge groups.

In the fundamental representation of $SU(2)_L$, the $SU(2)_L$ index symmetries of leptons, antileptons, and $W$ bosons are represented by the standard Young tableaux (SYT) as:
\begin{equation}
\label{eq:color_inside_symmetry}
l^{c_1} \sim \Yvcentermath1\young(\onec)
\quad\quad\quad
\bar{l}^{c_1} \sim \Yvcentermath1\young(\onec)
\quad\quad\quad
W^{c_1 c_2} \sim \Yvcentermath1\young(\onec \twoc)
\end{equation}
The set of color structures $\{T\}$ corresponds to the tensor product of the above tableaux into $SU(2)$ singlets. A more direct approach is to list all singlet SYTs $\{T'\}$ in the fundamental representation and project out the gauge factors $\{T\}$ satisfying the index symmetries of each particle using the Young operators in Eq.~(\ref{eq:color_inside_symmetry})~\cite{Dong:2022jru}:
\begin{equation}
\{T \} \sim \mathcal{Y}_{\text{\scriptsize\young(\onec)}} \times \mathcal{Y}_{\text{\scriptsize\young(\twoc)}} \times \mathcal{Y}_{\text{\scriptsize\young(\threec \fourc)}} \times \mathcal{Y}_{\text{\scriptsize\young(\fivec \sixc)}}\, \{T^\prime \} \equiv \mathcal{P} \cdot \{T^\prime \} \,.
\label{eq:numberbox}
\end{equation}
For example, consider the four-point scattering process \( l\!-\!\bar{l}\!-\!W_1\!-\!W_2 \). The singlet SYT corresponds to the tableau \( [3, 3] \), which has five $T'$. After projection, only two remain:
\begin{equation} 
\label{eq:Tbases}
\{T\}= \mathcal{P} \cdot \left\{
    \Yvcentermath1
    \young(\onec \twoc \threec,\fourc \fivec \sixc),\
    \young(\onec \threec \fourc,\twoc \fivec \sixc)
\right\}\,.
\end{equation}
As an example of index contraction, the color factor corresponding to the SYT is
\begin{equation}
\epsilon_{c_1 c_4} \epsilon_{c_2 c_5} \epsilon_{c_3 c_6} 
\,l^{c_1}\,\bar{l}^{c_2}\,W^{c_3 c_4}_1\,W^{c_5 c_6}_2 \,.
\label{eq:Tstructure}
\end{equation}
Since the symmetry of gauge indices is already encoded in the external field operators, we can directly use $T'$ as the color factor instead of strictly using $T$.

Once the $\{T\}$ basis is obtained, the matrix $M^T_{(12),(1,\ldots,n)}$ can be derived by applying permutation elements to the color indices of identical particles. For instance, for identical bosons $W_1$ and $W_2$, the permutation $(12)$ corresponds to swapping the color indices $3^c \leftrightarrow 5^c$ and $4^c \leftrightarrow 6^c$ in the SYT. 

Eq.~(\ref{eq:color_inside_symmetry}) implies that the permutation is closed under the set $\{T\}$, allowing us to perform the permutation first and then project:
\begin{equation}
\begin{aligned}
\label{eq:S3_symmetrize}
&M_{[r_n]}=\mathcal{Y}_{[r_n]}(x',y')\cdot P,\\
&x'=M^{T^\prime}_{(12)}\otimes M^\mathcal{L}_{(12)},\quad y'=M^{T^\prime}_{(123)}\otimes M^\mathcal{L}_{(123)}\,,
\end{aligned}
\end{equation}
where $M^{T^\prime}$ is the representation matrix in the $\{T'\}$ space, obtained by normalizing the Young tableau of each permutation $(12)T'$ into a linear combination of $T'$.

\section{Operators for Dark Photon and SM Fields}
\label{app:DP}
We list all operators involving the dark photon and the SM sector in the following tables. 
We define the vector dark photon as $X^{\mu}$ and its field strength as $X^{\pm}_{\mu\nu}$.
The extra $*$ mark after some types indicates that these types are invariant subspaces under complex conjugation. 

In this section we use the fundamental representations of $SU(2)_L,SU(3)_c$ to represent gauge bosons while other indices should be converted to the fundamental representation via generators $\lambda$ and Hodge duality $\epsilon$.
\begin{equation}
    G^{b_1 b_2 b_3}_{\mu \nu}\equiv G^{i}_{\mu \nu} \lambda_{i,\bar{b}_1}^{b_1} \epsilon^{\bar{b}_i b_2 b_3}
\end{equation}
where $i$ is the index of the adjoint representation. $b_{1,2,3}$ are the fundamental representation indices, and $\bar{b}_1$ is the anti-fundamental representation index. Indices in both superscript and subscript are summed over.

\begin{longtable}{cp{2cm}p{12cm}}
\caption{4pt dim-6 operator with interaction between Dark photon and SM sector}
\label{DP46} \\ \hline
Class & Type & Form\\\hline

\multirow{8}{*}{$V V V V $}&	\multirow{1}{*}{$ X X B B $}	&${X}_{\mu}^{} {X}_{\nu}^{} {B}_{\rho \xi}^{-\ } {B}_{\tau \zeta}^{-\ } g^{\nu\mu} \operatorname{Tr} \left( \sigma^{\tau\zeta} \sigma^{\rho\xi} \right)$	\\ \cline{2-3}
&	\multirow{1}{*}{$ X X G G $}	&$\epsilon_{b_{1} b_{3} b_{5} } \epsilon_{b_{2} b_{4} b_{6} } {X}_{\mu}^{} {X}_{\nu}^{} {G}_{\rho \xi}^{-\ b_{1} b_{2} b_{3} } {G}_{\tau \zeta}^{-\ b_{4} b_{5} b_{6} } g^{\nu\mu} \operatorname{Tr} \left( \sigma^{\tau\zeta} \sigma^{\rho\xi} \right)$	\\ \cline{2-3}
&	\multirow{1}{*}{$ X X W W $}	&$\epsilon_{a_{1} a_{3} } \epsilon_{a_{2} a_{4} } {X}_{\mu}^{} {X}_{\nu}^{} {W}_{\rho \xi}^{-\ a_{1} a_{2} } {W}_{\tau \zeta}^{-\ a_{3} a_{4} } g^{\nu\mu} \operatorname{Tr} \left( \sigma^{\tau\zeta} \sigma^{\rho\xi} \right)$	\\ \cline{2-3}
&	\multirow{1}{*}{$ X X B B ^{*} $}	&${X}_{\mu}^{} {X}_{\nu}^{} {B}_{\rho \xi}^{+\ } {B}_{\tau \zeta}^{-\ } \operatorname{Tr} \left( \sigma^{\tau\zeta} \sigma^{\nu} \bar{\sigma}^{\rho\xi} \bar{\sigma}^{\mu} \right)$	\\ \cline{2-3}
&	\multirow{2}{*}{$ X X X B $}	&${X}_{\mu \nu}^{+\ } {X}_{\rho}^{} {X}_{\xi}^{} {B}_{\tau \zeta}^{-\ } \operatorname{Tr} \left( \sigma^{\tau\zeta} \sigma^{\rho} \sigma^{\mu\nu} \bar{\sigma}^{\xi} \right)$	 \\ \cline{3-3}
&	 	&${X}_{\mu}^{} {X}_{\nu}^{} {X}_{\rho \xi}^{-\ } {B}_{\tau \zeta}^{-\ } g^{\nu\mu} \operatorname{Tr} \left( \sigma^{\tau\zeta} \sigma^{\rho\xi} \right)$	\\
\hline

\end{longtable}
\begin{longtable}{cp{2cm}p{12cm}}
\caption{5pt dim-6 operator with interaction between Dark photon and SM sector}
\label{DP56} \\ \hline
Class & Type & Form\\\hline

\multirow{2}{*}{$\phi \phi V V V $}&	\multirow{2}{*}{$ X X X H^\dagger H ^{*} $}	&$\epsilon_{a_{1} a_{2} } D_{\mu} {H}^{a_{2} } {H^\dagger}^{a_{1} } {X}_{\nu}^{} {X}_{\rho}^{} {X}_{\xi}^{} g^{\xi\mu} g^{\rho\nu}$	 \\ \cline{3-3}
&	 	&$\epsilon_{a_{1} a_{2} } D_{\mu} {H^\dagger}^{a_{1} } {H}^{a_{2} } {X}_{\nu}^{} {X}_{\rho}^{} {X}_{\xi}^{} g^{\xi\mu} g^{\rho\nu}$	\\ \hline
\multirow{5}{*}{$\psi \psi V V V $}&	\multirow{1}{*}{$ X X X \bar{d} d ^{*} $}	&$\epsilon_{b_{1} b_{2} b_{3} } \left( \left( {d}_{b_{1} ,R}^{} \right) \bar{\sigma}^{\mu} \left( {d}_{b_{2} ,R}^{\dagger\ b_{3} } \right) \right) {X}_{\nu}^{} {X}_{\rho}^{} {X}_{\mu}^{} g^{\rho\nu}$	\\ \cline{2-3}
&	\multirow{1}{*}{$ X X X \bar{e} e ^{*} $}	&$\left( \left( {e}_{R}^{} \right) \bar{\sigma}^{\mu} \left( {e}_{R}^{\dagger\ } \right) \right) {X}_{\nu}^{} {X}_{\rho}^{} {X}_{\mu}^{} g^{\rho\nu}$	\\ \cline{2-3}
&	\multirow{1}{*}{$ X X X \bar{L} L ^{*} $}	&$\epsilon_{a_{1} a_{2} } \left( \left( {L}_{a_{1} ,L}^{\dagger\ } \right) \bar{\sigma}^{\mu} \left( {L}_{a_{2} ,L}^{} \right) \right) {X}_{\nu}^{} {X}_{\rho}^{} {X}_{\mu}^{} g^{\rho\nu}$	\\ \cline{2-3}
&	\multirow{1}{*}{$ X X X \bar{q} q ^{*} $}	&$\epsilon_{a_{1} a_{2} } \epsilon_{b_{1} b_{2} b_{3} } \left( \left( {q}_{b_{2} ,L}^{\dagger\ b_{3} a_{1} } \right) \bar{\sigma}^{\mu} \left( {q}_{b_{1} ,L}^{a_{2} } \right) \right) {X}_{\nu}^{} {X}_{\rho}^{} {X}_{\mu}^{} g^{\rho\nu}$	\\
\hline

\end{longtable}
\begin{longtable}{cp{2cm}p{12cm}}
\caption{6pt dim-6 operator with interaction between Dark photon and SM sector}
\label{DP66} \\ \hline
Class & Type & Form\\\hline

\multirow{1}{*}{$\phi \phi V V V V $}&	\multirow{1}{*}{$ X X X X H^\dagger H ^{*} $}	&$\epsilon_{a_{1} a_{2} } {H^\dagger}^{a_{1} } {H}^{a_{2} } {X}_{\mu}^{} {X}_{\nu}^{} {X}_{\rho}^{} {X}_{\xi}^{} g^{\xi\rho} g^{\nu\mu}$	\\ \hline
\multirow{1}{*}{$\phi \phi \phi \phi V V $}&	\multirow{1}{*}{$ X X H^\dagger H^\dagger H H ^{*} $}	&$\epsilon_{a_{1} a_{3} } \epsilon_{a_{2} a_{4} } {H^\dagger}^{a_{1} } {H^\dagger}^{a_{2} } {H}^{a_{3} } {H}^{a_{4} } {X}_{\mu}^{} {X}_{\nu}^{} g^{\nu\mu}$	\\ \hline

\end{longtable}
\begin{longtable}{cp{2cm}p{12cm}}
\caption{4pt dim-8 operator with interaction between Dark photon and SM sector}
\label{DP48} \\ \hline
Class & Type & Form\\\hline

\multirow{63}{*}{$V V V V $}&	\multirow{1}{*}{$ X B B B $}	&${X}_{\mu \nu}^{-\ } {B}_{\rho \xi}^{-\ } {B}_{\tau \zeta}^{-\ } {B}_{\eta \theta}^{-\ } \operatorname{Tr} \left( \sigma^{\eta\theta} \sigma^{\tau\zeta} \right) \operatorname{Tr} \left( \sigma^{\rho\xi} \sigma^{\mu\nu} \right)$	\\ \cline{2-3}
&	\multirow{3}{*}{$ X G G B $}	&$\epsilon_{b_{1} b_{3} b_{5} } \epsilon_{b_{2} b_{4} b_{6} } {X}_{\mu \nu}^{-\ } {G}_{\rho \xi}^{-\ b_{1} b_{2} b_{3} } {G}_{\tau \zeta}^{-\ b_{4} b_{5} b_{6} } {B}_{\eta \theta}^{-\ } \operatorname{Tr} \left( \sigma^{\eta\theta} \sigma^{\tau\zeta} \right) \operatorname{Tr} \left( \sigma^{\rho\xi} \sigma^{\mu\nu} \right)$	 \\ \cline{3-3}
&	 	&$\epsilon_{b_{1} b_{3} b_{5} } \epsilon_{b_{2} b_{4} b_{6} } {X}_{\mu \nu}^{-\ } {G}_{\rho \xi}^{-\ b_{1} b_{2} b_{3} } {G}_{\tau \zeta}^{-\ b_{4} b_{5} b_{6} } {B}_{\eta \theta}^{-\ } \operatorname{Tr} \left( \sigma^{\eta\theta} \bar{\sigma}^{\rho\xi} \sigma^{\mu\nu} \sigma^{\tau\zeta} \right)$	 \\ \cline{3-3}
&	 	&$\epsilon_{b_{1} b_{3} b_{5} } \epsilon_{b_{2} b_{4} b_{6} } D_{\mu} {G}_{\nu \rho}^{-\ b_{4} b_{5} b_{6} } {X}_{\xi}^{} {G}_{\tau \zeta}^{-\ b_{1} b_{2} b_{3} } {B}_{\eta \theta}^{-\ } \operatorname{Tr} \left( \sigma^{\eta\theta} \sigma^{\nu\rho} \right) \operatorname{Tr} \left( \sigma^{\tau\zeta} \sigma^{\mu} \bar{\sigma}^{\xi} \right)$	\\ \cline{2-3}
&	\multirow{3}{*}{$ X W W B $}	&$\epsilon_{a_{1} a_{3} } \epsilon_{a_{2} a_{4} } {X}_{\mu \nu}^{-\ } {W}_{\rho \xi}^{-\ a_{1} a_{2} } {W}_{\tau \zeta}^{-\ a_{3} a_{4} } {B}_{\eta \theta}^{-\ } \operatorname{Tr} \left( \sigma^{\eta\theta} \sigma^{\tau\zeta} \right) \operatorname{Tr} \left( \sigma^{\rho\xi} \sigma^{\mu\nu} \right)$	 \\
\hline

\end{longtable}
\begin{longtable}{cp{2cm}p{12cm}}
\caption{5pt dim-8 operator with interaction between Dark photon and SM sector}
\label{DP58} \\ \hline
Class & Type & Form\\\hline

\multirow{35}{*}{$V V V V V $}&	\multirow{1}{*}{$ X X G G G $}	&$\epsilon_{b_{1} b_{4} b_{6} } \epsilon_{b_{2} b_{5} b_{8} } \epsilon_{b_{3} b_{7} b_{9} } {X}_{\mu}^{} {X}_{\nu}^{} {G}_{\rho \xi}^{-\ b_{1} b_{2} b_{3} } {G}_{\tau \zeta}^{-\ b_{4} b_{5} b_{6} } {G}_{\eta \theta}^{-\ b_{7} b_{8} b_{9} } g^{\nu\mu} \operatorname{Tr} \left( \sigma^{\eta\theta} \bar{\sigma}^{\rho\xi} \sigma^{\tau\zeta} \right)$	\\ \cline{2-3}
&	\multirow{1}{*}{$ X X W W W $}	&$\epsilon_{a_{1} a_{4} } \epsilon_{a_{2} a_{5} } \epsilon_{a_{3} a_{6} } {X}_{\mu}^{} {X}_{\nu}^{} {W}_{\rho \xi}^{-\ a_{1} a_{2} } {W}_{\tau \zeta}^{-\ a_{3} a_{4} } {W}_{\eta \theta}^{-\ a_{5} a_{6} } g^{\nu\mu} \operatorname{Tr} \left( \sigma^{\eta\theta} \bar{\sigma}^{\rho\xi} \sigma^{\tau\zeta} \right)$	\\ \cline{2-3}
&	\multirow{2}{*}{$ X X X B B $}	&$D_{\mu} {B}_{\nu \rho}^{-\ } {X}_{\xi}^{} {X}_{\tau}^{} {X}_{\zeta}^{} {B}_{\eta \theta}^{-\ } g^{\zeta\mu} g^{\tau\xi} \operatorname{Tr} \left( \sigma^{\nu\rho} \sigma^{\eta\theta} \right)$	 \\ \cline{3-3}
&	 	&$D_{\mu} {X}_{\nu}^{} {X}_{\rho}^{} {X}_{\xi}^{} {B}_{\tau \zeta}^{-\ } {B}_{\eta \theta}^{-\ } \operatorname{Tr} \left( \sigma^{\eta\theta} \sigma^{\mu} \bar{\sigma}^{\xi} \sigma^{\tau\zeta} \sigma^{\rho} \bar{\sigma}^{\nu} \right)$	\\ \cline{2-3}
&	\multirow{1}{*}{$ X X G G B $}	&$\epsilon_{b_{1} b_{3} b_{5} } \epsilon_{b_{2} b_{4} b_{6} } {X}_{\mu}^{} {X}_{\nu}^{} {G}_{\rho \xi}^{+\ b_{1} b_{2} b_{3} } {G}_{\tau \zeta}^{-\ b_{4} b_{5} b_{6} } {B}_{\eta \theta}^{-\ } \operatorname{Tr} \left( \sigma^{\eta\theta} \sigma^{\nu} \bar{\sigma}^{\rho\xi} \bar{\sigma}^{\mu} \sigma^{\tau\zeta} \right)$	\\
\hline

\end{longtable}
\begin{longtable}{cp{2cm}p{12cm}}
\caption{6pt dim-8 operator with interaction between Dark photon and SM sector}
\label{DP68} \\ \hline
Class & Type & Form\\\hline

\multirow{8}{*}{$V V V V V V $}&	\multirow{1}{*}{$ X X X X B B $}	&${X}_{\mu}^{} {X}_{\nu}^{} {X}_{\rho}^{} {X}_{\xi}^{} {B}_{\tau \zeta}^{-\ } {B}_{\eta \theta}^{-\ } g^{\xi\rho} g^{\nu\mu} \operatorname{Tr} \left( \sigma^{\eta\theta} \sigma^{\tau\zeta} \right)$	\\ \cline{2-3}
&	\multirow{1}{*}{$ X X X X G G $}	&$\epsilon_{b_{1} b_{3} b_{5} } \epsilon_{b_{2} b_{4} b_{6} } {X}_{\mu}^{} {X}_{\nu}^{} {X}_{\rho}^{} {X}_{\xi}^{} {G}_{\tau \zeta}^{-\ b_{1} b_{2} b_{3} } {G}_{\eta \theta}^{-\ b_{4} b_{5} b_{6} } g^{\xi\rho} g^{\nu\mu} \operatorname{Tr} \left( \sigma^{\eta\theta} \sigma^{\tau\zeta} \right)$	\\ \cline{2-3}
&	\multirow{1}{*}{$ X X X X W W $}	&$\epsilon_{a_{1} a_{3} } \epsilon_{a_{2} a_{4} } {X}_{\mu}^{} {X}_{\nu}^{} {X}_{\rho}^{} {X}_{\xi}^{} {W}_{\tau \zeta}^{-\ a_{1} a_{2} } {W}_{\eta \theta}^{-\ a_{3} a_{4} } g^{\xi\rho} g^{\nu\mu} \operatorname{Tr} \left( \sigma^{\eta\theta} \sigma^{\tau\zeta} \right)$	\\ \cline{2-3}
&	\multirow{1}{*}{$ X X X X B B ^{*} $}	&${X}_{\mu}^{} {X}_{\nu}^{} {X}_{\rho}^{} {X}_{\xi}^{} {B}_{\tau \zeta}^{+\ } {B}_{\eta \theta}^{-\ } g^{\nu\mu} \operatorname{Tr} \left( \sigma^{\eta\theta} \sigma^{\xi} \bar{\sigma}^{\tau\zeta} \bar{\sigma}^{\rho} \right)$	\\ \cline{2-3}
&	\multirow{2}{*}{$ X X X X X B $}	&${X}_{\mu \nu}^{+\ } {X}_{\rho}^{} {X}_{\xi}^{} {X}_{\tau}^{} {X}_{\zeta}^{} {B}_{\eta \theta}^{-\ } \operatorname{Tr} \left( \sigma^{\eta\theta} \sigma^{\tau} \bar{\sigma}^{\mu\nu} \bar{\sigma}^{\rho} \sigma^{\xi} \bar{\sigma}^{\zeta} \right)$	 \\ \cline{3-3}
&	 	&${X}_{\mu}^{} {X}_{\nu}^{} {X}_{\rho}^{} {X}_{\xi}^{} {X}_{\tau \zeta}^{-\ } {B}_{\eta \theta}^{-\ } g^{\xi\rho} g^{\nu\mu} \operatorname{Tr} \left( \sigma^{\eta\theta} \sigma^{\tau\zeta} \right)$	\\
\hline

\end{longtable}
\section{Operators for Gravitino Like Particles and SM Fields}
\label{app:Gravitino}
We list all operators involving the spin-$\frac{3}{2}$ gravitino like particle and the SM sector in the following tables. The gravitino like particle is defined as $\psi^{\mu}$. The extra $*$ mark after some types indicates that these types are invariant subspaces under complex conjugation.

\begin{longtable}{cp{2cm}p{12cm}}
\caption{4pt dim-6 operator with interaction between Gravitino and SM sector}
\label{Grav46} \\ \hline
Class & Type & Form\\\hline

\multirow{2}{*}{$\phi \psi \psi^{\mu}V $}&	\multirow{1}{*}{$\psi L H B$}	&$\epsilon_{a_{1} a_{2} } \left( \left( {L}_{a_{1} ,L}^{\dagger\ } \right) \bar{\sigma}^{\mu} \sigma^{\nu\rho} \left( {\psi}_{\mu,L}^{} \right) \right) {H}^{a_{2} } {B}_{\nu \rho}^{-\ }$	\\ \cline{2-3}
&	\multirow{1}{*}{$\psi W L H$}	&$\epsilon_{a_{1} a_{3} } \epsilon_{a_{2} a_{4} } \left( \left( {L}_{a_{1} ,L}^{\dagger\ } \right) \bar{\sigma}^{\mu} \sigma^{\nu\rho} \left( {\psi}_{\mu,L}^{} \right) \right) {H}^{a_{2} } {W}_{\nu \rho}^{-\ a_{3} a_{4} }$	\\ \hline
\multirow{10}{*}{$\psi \psi \psi^{\mu}\psi^{\mu}$}&	\multirow{2}{*}{$\psi \psi d d ^{*}$}	&$\epsilon_{b_{1} b_{2} b_{3} } \left( \left( {\psi}_{\mu,L}^{\dagger\ } \right) \bar{\sigma}^{\nu} \left( {d}_{b_{2} ,R}^{\dagger\ b_{3} } \right) \right) \left( \left( {\psi}_{\nu,L}^{} \right) \sigma^{\mu} \left( {d}_{b_{1} ,R}^{} \right) \right)$	 \\ \cline{3-3}
&	 	&$\epsilon_{b_{1} b_{2} b_{3} } \left( \left( {\psi}_{\mu,R}^{\dagger\ } \right) \sigma^{\nu} \left( {d}_{b_{1} ,R}^{} \right) \right) \left( \left( {\psi}_{\nu,R}^{} \right) \bar{\sigma}^{\mu} \left( {d}_{b_{2} ,R}^{\dagger\ b_{3} } \right) \right)$	\\ \cline{2-3}
&	\multirow{2}{*}{$\psi \psi e e ^{*}$}	&$\left( \left( {\psi}_{\mu,L}^{\dagger\ } \right) \bar{\sigma}^{\nu} \left( {e}_{R}^{\dagger\ } \right) \right) \left( \left( {\psi}_{\nu,L}^{} \right) \sigma^{\mu} \left( {e}_{R}^{} \right) \right)$	 \\ \cline{3-3}
&	 	&$\left( \left( {\psi}_{\mu,R}^{\dagger\ } \right) \sigma^{\nu} \left( {e}_{R}^{} \right) \right) \left( \left( {\psi}_{\nu,R}^{} \right) \bar{\sigma}^{\mu} \left( {e}_{R}^{\dagger\ } \right) \right)$	\\
\hline

\end{longtable}
\begin{longtable}{cp{2cm}p{10cm}}
\caption{4pt dim-7 operator with interaction between Gravitino and SM sector}
\label{Grav47} \\ \hline
Class & Type & Form\\\hline

\multirow{24}{*}{$\psi^{\mu}\psi^{\mu}V V $}&	\multirow{3}{*}{$\psi \psi B B$}	&$\left( \left( {\psi}_{\mu,L}^{\dagger\ } \right) \left( {\psi}_{\nu,R}^{} \right) \right) {B}_{\rho \xi}^{-\ } {B}_{\tau \zeta}^{-\ } g^{\nu\mu} \operatorname{Tr} \left( \sigma^{\tau\zeta} \sigma^{\rho\xi} \right)$	 \\ \cline{3-3}
&	 	&$\left( \left( {\psi}_{\mu,R}^{\dagger\ } \right) \left( {\psi}_{\nu,L}^{} \right) \right) {B}_{\rho \xi}^{-\ } {B}_{\tau \zeta}^{-\ } g^{\nu\mu} \operatorname{Tr} \left( \sigma^{\tau\zeta} \sigma^{\rho\xi} \right)$	 \\ \cline{3-3}
&	 	&$\left( \left( {\psi}_{\mu,R}^{\dagger\ } \right) \bar{\sigma}^{\nu\rho} \sigma^{\mu} \bar{\sigma}^{\xi} \sigma^{\tau\zeta} \left( {\psi}_{\xi,L}^{} \right) \right) {B}_{\nu \rho}^{-\ } {B}_{\tau \zeta}^{-\ }$	\\ \cline{2-3}
&	\multirow{3}{*}{$\psi \psi G G$}	&$\epsilon_{b_{1} b_{3} b_{5} } \epsilon_{b_{2} b_{4} b_{6} } \left( \left( {\psi}_{\mu,L}^{\dagger\ } \right) \left( {\psi}_{\nu,R}^{} \right) \right) {G}_{\rho \xi}^{-\ b_{1} b_{2} b_{3} } {G}_{\tau \zeta}^{-\ b_{4} b_{5} b_{6} } g^{\nu\mu} \operatorname{Tr} \left( \sigma^{\tau\zeta} \sigma^{\rho\xi} \right)$	 \\ \cline{3-3}
&	 	&$\epsilon_{b_{1} b_{3} b_{5} } \epsilon_{b_{2} b_{4} b_{6} } \left( \left( {\psi}_{\mu,R}^{\dagger\ } \right) \left( {\psi}_{\nu,L}^{} \right) \right) {G}_{\rho \xi}^{-\ b_{1} b_{2} b_{3} } {G}_{\tau \zeta}^{-\ b_{4} b_{5} b_{6} } g^{\nu\mu} \operatorname{Tr} \left( \sigma^{\tau\zeta} \sigma^{\rho\xi} \right)$	 \\ \cline{3-3}
&	 	&$\epsilon_{b_{1} b_{3} b_{5} } \epsilon_{b_{2} b_{4} b_{6} } \left( \left( {\psi}_{\mu,R}^{\dagger\ } \right) \bar{\sigma}^{\nu\rho} \sigma^{\mu} \bar{\sigma}^{\xi} \sigma^{\tau\zeta} \left( {\psi}_{\xi,L}^{} \right) \right) {G}_{\nu \rho}^{-\ b_{1} b_{2} b_{3} } {G}_{\tau \zeta}^{-\ b_{4} b_{5} b_{6} }$	\\
\hline

\end{longtable}
\begin{longtable}{cp{2cm}p{12cm}}
\caption{5pt dim-7 operator with interaction between Gravitino and SM sector}
\label{Grav57} \\ \hline
Class & Type & Form\\\hline

\multirow{1}{*}{$\phi \phi \psi \psi^{\mu}V $}&	\multirow{1}{*}{$\psi W H H e$}	&$\epsilon_{a_{1} a_{3} } \epsilon_{a_{2} a_{4} } \left( \left( {e}_{R}^{\dagger\ } \right) \sigma^{\mu} \bar{\sigma}^{\nu\rho} \left( {\psi}_{\mu,R}^{} \right) \right) {H}^{a_{1} } {H}^{a_{2} } {W}_{\nu \rho}^{+\ a_{3} a_{4} }$	\\ \hline
\multirow{8}{*}{$\phi \phi \psi^{\mu}\psi^{\mu}V $}&	\multirow{2}{*}{$\psi \psi H^\dagger H B$}	&$\epsilon_{a_{1} a_{2} } \left( \left( {\psi}_{\mu,L}^{\dagger\ } \right) \left( {\psi}_{\nu,R}^{} \right) \right) {H^\dagger}^{a_{1} } {H}^{a_{2} } {B}_{\rho \xi}^{-\ } \operatorname{Tr} \left( \bar{\sigma}^{\nu} \bar{\sigma}^{\rho\xi} \sigma^{\mu} \right)$	 \\ \cline{3-3}
&	 	&$\epsilon_{a_{1} a_{2} } \left( \left( {\psi}_{\mu,R}^{\dagger\ } \right) \sigma^{\nu\rho} \left( {\psi}_{\xi,L}^{} \right) \right) {H^\dagger}^{a_{1} } {H}^{a_{2} } {B}_{\nu \rho}^{-\ } g^{\xi\mu}$	\\ \cline{2-3}
&	\multirow{2}{*}{$\psi \psi W H^\dagger H$}	&$\epsilon_{a_{1} a_{3} } \epsilon_{a_{2} a_{4} } \left( \left( {\psi}_{\mu,L}^{\dagger\ } \right) \left( {\psi}_{\nu,R}^{} \right) \right) {H^\dagger}^{a_{1} } {H}^{a_{2} } {W}_{\rho \xi}^{-\ a_{3} a_{4} } \operatorname{Tr} \left( \bar{\sigma}^{\nu} \bar{\sigma}^{\rho\xi} \sigma^{\mu} \right)$	 \\ \cline{3-3}
&	 	&$\epsilon_{a_{1} a_{3} } \epsilon_{a_{2} a_{4} } \left( \left( {\psi}_{\mu,R}^{\dagger\ } \right) \sigma^{\nu\rho} \left( {\psi}_{\xi,L}^{} \right) \right) {H^\dagger}^{a_{1} } {H}^{a_{2} } {W}_{\nu \rho}^{-\ a_{3} a_{4} } g^{\xi\mu}$	\\ \cline{2-3}
&	\multirow{2}{*}{$\psi \psi H^\dagger H B$}	&$\epsilon_{a_{1} a_{2} } \left( \left( {\psi}_{\mu,L}^{\dagger\ } \right) \bar{\sigma}^{\nu\rho} \left( {\psi}_{\xi,R}^{} \right) \right) {H^\dagger}^{a_{1} } {H}^{a_{2} } {B}_{\nu \rho}^{+\ } g^{\xi\mu}$	 \\
\hline

\end{longtable}
\begin{longtable}{cp{2cm}p{12cm}}
\caption{6pt dim-7 operator with interaction between Gravitino and SM sector}
\label{Grav67} \\ \hline
Class & Type & Form\\\hline
\multirow{2}{*}{$\phi \phi \phi \phi \psi^{\mu}\psi^{\mu}$}&	\multirow{2}{*}{$\psi \psi H^\dagger H^\dagger H H ^{*}$}	&$\epsilon_{a_{1} a_{3} } \epsilon_{a_{2} a_{4} } \left( \left( {\psi}_{\mu,L}^{\dagger\ } \right) \left( {\psi}_{\nu,R}^{} \right) \right) {H^\dagger}^{a_{1} } {H^\dagger}^{a_{2} } {H}^{a_{3} } {H}^{a_{4} } g^{\nu\mu}$	 \\ \cline{3-3}
&	 	&$\epsilon_{a_{1} a_{3} } \epsilon_{a_{2} a_{4} } \left( \left( {\psi}_{\mu,R}^{\dagger\ } \right) \left( {\psi}_{\nu,L}^{} \right) \right) {H^\dagger}^{a_{1} } {H^\dagger}^{a_{2} } {H}^{a_{3} } {H}^{a_{4} } g^{\nu\mu}$	\\ \hline

\end{longtable}
\begin{longtable}{cp{2cm}p{12cm}}
\caption{4pt dim-8 operator with interaction between Gravitino and SM sector}
\label{Grav57} \\ \hline
Class & Type & Form\\\hline

\multirow{36}{*}{$\psi^{\mu}\psi^{\mu}V V $}&	\multirow{2}{*}{$\psi \psi B B$}	&$\left( \left( {\psi}_{\mu,L}^{\dagger\ } \right) \bar{\sigma}^{\nu} \sigma^{\rho\xi} \left( D_{\tau} {\psi}_{\nu,L}^{} \right) \right) {B}_{\zeta \eta}^{-\ } {B}_{\rho \xi}^{-\ } \operatorname{Tr} \left( \sigma^{\mu} \bar{\sigma}^{\tau} \sigma^{\zeta\eta} \right)$	 \\ \cline{3-3}
&	 	&$\left( \left( {\psi}_{\mu,R}^{\dagger\ } \right) \sigma^{\nu\rho} \sigma^{\xi\tau} \sigma^{\zeta} \bar{\sigma}^{\eta} \bar{\sigma}^{\mu} \left( {\psi}_{\zeta,R}^{} \right) \right) D_{\eta} {B}_{\xi \tau}^{-\ } {B}_{\nu \rho}^{-\ }$	\\ \cline{2-3}
&	\multirow{2}{*}{$\psi \psi G G$}	&$\epsilon_{b_{1} b_{3} b_{5} } \epsilon_{b_{2} b_{4} b_{6} } \left( \left( {\psi}_{\mu,L}^{\dagger\ } \right) \bar{\sigma}^{\nu} \sigma^{\rho\xi} \left( D_{\tau} {\psi}_{\nu,L}^{} \right) \right) {G}_{\zeta \eta}^{-\ b_{1} b_{2} b_{3} } {G}_{\rho \xi}^{-\ b_{4} b_{5} b_{6} } \operatorname{Tr} \left( \sigma^{\mu} \bar{\sigma}^{\tau} \sigma^{\zeta\eta} \right)$	 \\ \cline{3-3}
&	 	&$\epsilon_{b_{1} b_{3} b_{5} } \epsilon_{b_{2} b_{4} b_{6} } \left( \left( {\psi}_{\mu,R}^{\dagger\ } \right) \sigma^{\nu\rho} \sigma^{\xi\tau} \sigma^{\zeta} \bar{\sigma}^{\eta} \bar{\sigma}^{\mu} \left( {\psi}_{\zeta,R}^{} \right) \right) D_{\eta} {G}_{\xi \tau}^{-\ b_{4} b_{5} b_{6} } {G}_{\nu \rho}^{-\ b_{1} b_{2} b_{3} }$	\\ \cline{2-3}
&	\multirow{2}{*}{$\psi \psi W W$}	&$\epsilon_{a_{1} a_{3} } \epsilon_{a_{2} a_{4} } \left( \left( {\psi}_{\mu,L}^{\dagger\ } \right) \bar{\sigma}^{\nu} \sigma^{\rho\xi} \left( D_{\tau} {\psi}_{\nu,L}^{} \right) \right) {W}_{\zeta \eta}^{-\ a_{1} a_{2} } {W}_{\rho \xi}^{-\ a_{3} a_{4} } \operatorname{Tr} \left( \sigma^{\mu} \bar{\sigma}^{\tau} \sigma^{\zeta\eta} \right)$	 \\
\hline

\end{longtable}
\begin{longtable}{cp{2cm}p{12cm}}
\caption{5pt dim-8 operator with interaction between Gravitino and SM sector}
\label{Grav58} \\ \hline
Class & Type & Form\\\hline

\multirow{8}{*}{$\phi \psi \psi^{\mu}V V $}&	\multirow{1}{*}{$\psi W L H B$}	&$\epsilon_{a_{1} a_{3} } \epsilon_{a_{2} a_{4} } \left( \left( {L}_{a_{1} ,L}^{\dagger\ } \right) \bar{\sigma}^{\mu} \sigma^{\nu\rho} \sigma^{\xi\tau} \left( {\psi}_{\mu,L}^{} \right) \right) {H}^{a_{2} } {W}_{\nu \rho}^{-\ a_{3} a_{4} } {B}_{\xi \tau}^{-\ }$	\\ \cline{2-3}
&	\multirow{1}{*}{$\psi W W L H$}	&$\epsilon_{a_{1} a_{4} } \epsilon_{a_{2} a_{5} } \epsilon_{a_{3} a_{6} } \left( \left( {L}_{a_{1} ,L}^{\dagger\ } \right) \bar{\sigma}^{\mu} \sigma^{\nu\rho} \sigma^{\xi\tau} \left( {\psi}_{\mu,L}^{} \right) \right) {H}^{a_{2} } {W}_{\nu \rho}^{-\ a_{3} a_{4} } {W}_{\xi \tau}^{-\ a_{5} a_{6} }$	\\ \cline{2-3}
&	\multirow{1}{*}{$\psi L H B B$}	&$\epsilon_{a_{1} a_{2} } \left( \left( {L}_{a_{1} ,L}^{\dagger\ } \right) \sigma^{\mu\nu} \bar{\sigma}^{\rho} \sigma^{\xi\tau} \left( {\psi}_{\rho,L}^{} \right) \right) {H}^{a_{2} } {B}_{\mu \nu}^{+\ } {B}_{\xi \tau}^{-\ }$	\\ \cline{2-3}
&	\multirow{1}{*}{$\psi W L H B$}	&$\epsilon_{a_{1} a_{3} } \epsilon_{a_{2} a_{4} } \left( \left( {L}_{a_{1} ,L}^{\dagger\ } \right) \sigma^{\mu\nu} \bar{\sigma}^{\rho} \sigma^{\xi\tau} \left( {\psi}_{\rho,L}^{} \right) \right) {H}^{a_{2} } {W}_{\mu \nu}^{+\ a_{3} a_{4} } {B}_{\xi \tau}^{-\ }$	\\ \cline{2-3}
&	\multirow{1}{*}{$\psi W L H B$}	&$\epsilon_{a_{1} a_{3} } \epsilon_{a_{2} a_{4} } \left( \left( {L}_{a_{1} ,L}^{\dagger\ } \right) \sigma^{\mu\nu} \bar{\sigma}^{\rho} \sigma^{\xi\tau} \left( {\psi}_{\rho,L}^{} \right) \right) {H}^{a_{2} } {W}_{\xi \tau}^{-\ a_{3} a_{4} } {B}_{\mu \nu}^{+\ }$	\\
\hline

\end{longtable}
\begin{longtable}{cp{2cm}p{12cm}}
\caption{6pt dim-8 operator with interaction between Gravitino and SM sector}
\label{Grav68} \\ \hline
Class & Type & Form?\\\hline
\multirow{3}{*}{$\phi \phi \phi \psi \psi^{\mu}V $}&	\multirow{1}{*}{$\psi L H^\dagger H H B$}	&$\epsilon_{a_{1} a_{3} } \epsilon_{a_{2} a_{4} } \left( \left( {L}_{a_{1} ,L}^{\dagger\ } \right) \bar{\sigma}^{\mu} \sigma^{\nu\rho} \left( {\psi}_{\mu,L}^{} \right) \right) {H^\dagger}^{a_{2} } {H}^{a_{3} } {H}^{a_{4} } {B}_{\nu \rho}^{-\ }$	\\ \cline{2-3}
&	\multirow{2}{*}{$\psi W L H^\dagger H H$}	&$\epsilon_{a_{1} a_{4} } \epsilon_{a_{2} a_{5} } \epsilon_{a_{3} a_{6} } \left( \left( {L}_{a_{1} ,L}^{\dagger\ } \right) \bar{\sigma}^{\mu} \sigma^{\nu\rho} \left( {\psi}_{\mu,L}^{} \right) \right) {H^\dagger}^{a_{2} } {H}^{a_{3} } {H}^{a_{4} } {W}_{\nu \rho}^{-\ a_{5} a_{6} }$	 \\ \cline{3-3}
&	 	&$\epsilon_{a_{1} a_{3} } \epsilon_{a_{2} a_{5} } \epsilon_{a_{4} a_{6} } \left( \left( {L}_{a_{1} ,L}^{\dagger\ } \right) \bar{\sigma}^{\mu} \sigma^{\nu\rho} \left( {\psi}_{\mu,L}^{} \right) \right) {H^\dagger}^{a_{2} } {H}^{a_{3} } {H}^{a_{4} } {W}_{\nu \rho}^{-\ a_{5} a_{6} }$	\\ \hline
\multirow{3}{*}{$\phi \phi \psi \psi \psi \psi^{\mu}$}&	\multirow{1}{*}{$\psi u H^\dagger H e d$}	&$\epsilon_{a_{1} a_{2} } \epsilon_{b_{1} b_{2} b_{3} } \left( \left( {d}_{b_{1} ,R}^{} \right) \bar{\sigma}^{\mu} \left( {u}_{b_{2} ,R}^{\dagger\ b_{3} } \right) \right) \left( \left( {\psi}_{\mu,L}^{} \right) \left( {e}_{R}^{\dagger\ } \right) \right) {H^\dagger}^{a_{1} } {H}^{a_{2} }$	\\ \cline{2-3}
&	\multirow{1}{*}{$\psi L L H H e$}	&$\epsilon_{a_{1} a_{3} } \epsilon_{a_{2} a_{4} } \left( \left( {L}_{a_{1} ,L}^{\dagger\ } \right) \bar{\sigma}^{\mu} \left( {L}_{a_{2} ,L}^{} \right) \right) \left( \left( {\psi}_{\mu,L}^{} \right) \left( {e}_{R}^{\dagger\ } \right) \right) {H}^{a_{3} } {H}^{a_{4} }$	\\
\hline

\end{longtable}

\bibliographystyle{JHEP}
\bibliography{references.bib}

\end{document}